\newcommand{\be}{\begin{equation}}
\newcommand{\ee}{\end{equation}}
\newcommand{\ba}{\begin{eqnarray}}
\newcommand{\ea}{\end{eqnarray}}
\newcommand{\bc}{\begin{center}}
\newcommand{\ec}{\end{center}}

\newcommand{\bay}{\begin{array}{rcl}}
\newcommand{\eay}{\end{array}}
\newcommand{\dis}{\displaystyle}
\newcommand{\text}{\textstyle}
\def\RN{Reis\-sner-Nord\-str\"{o}m }
\documentstyle[aps,manuscript,epsfig,floats]{revtex}
\begin{document}
\baselineskip 18pt 
\rightline{INFNCT/03-00}
\rightline{MZ-TH/00-04}

\begin{center}
\large{{\bf Renormalization group improved black hole spacetimes}}

\ \\
\ \\

\normalsize{Alfio Bonanno\footnote{\tt abo@sunct.ct.astro.it}}

\small{\em Osservatorio Astrofisico,
Viale Andrea Doria 6, 95125 Catania, Italy\\
INFN Sezione di Catania, Corso Italia 57, 95100 Catania, Italy\\}

\ \\
\ \\

\normalsize{Martin Reuter\footnote{\tt reuter@thep.physik.uni-mainz.de}}

\small{\em Institut f\"ur Physik, Universit\"at Mainz\\
Staudingerweg 7, 55099 Mainz, Germany\\}

\end{center}

\begin{abstract}
We study the quantum gravitational effects in spherically symmetric
black hole spacetimes. The effective quantum spacetime felt by a point-like
test mass is constructed by ``renormalization group improving''  the  Schwarzschild metric.
The key ingredient is the running Newton constant which is obtained from the
exact evolution equation for the effective average action. The conformal structure of the 
quantum spacetime depends on its ADM-mass $M$ and it is similar to that of the classical
\RN black hole. For $M$ larger than, equal to, and smaller than a certain critical mass $M_{\rm cr}$
the spacetime has two, one and no horizon(s), respectively. Its Hawking temperature,
specific heat capacity and entropy are  computed as a function of $M$. It is argued that the 
black hole evaporation stops when $M$ approaches $M_{\rm cr}$ which is of the order 
of the Planck mass.  In this manner a ``cold'' soliton-like remnant with the near-horizon 
geometry of $AdS_2\times S^2$ is formed. As a consequence of the quantum effects, the 
classical singularity at $r=0$ is either removed completely or it is at least much milder than 
classically; in the first case the quantum spacetime has
a smooth de Sitter core which would be in accord with the cosmic censorship hypothesis
even if $M<M_{\rm cr}$. 
\end{abstract}

\draft
\pacs{97.60.Lf, 11.10.Hi, 04.60.-m}
\narrowtext

\renewcommand{\theequation}{1.\arabic{equation}}
\setcounter{equation}{0}

\section{introduction}
The Schwarzschild spacetime is the unique spherically symmetric vacuum solution
of Einstein's equations. Understanding the dynamics of this spacetime when quantum 
effects of the geometry are switched on has always been one of the most 
challenging issues from the theoretical point of view. It is in fact very plausible that 
those effects will play a key role in the very late stages of the 
gravitational collapse as well as during the evaporation process of a Planck size 
black hole.

According to the standard semiclassical scenario, a black hole of mass $M$ 
emits Hawking radiation at a temperature which is inversely proportional to $M$.
During this process, in addition to the radiation of energy to infinity,  
a negative-energy flux through the horizon is produced. 
Thereby the mass of the black hole is lowered and the temperature is increased.  
It is an open question whether this process continues until the entire mass of the black 
hole has been converted to radiation or whether it stops when the temperature is close 
to the Planck temperature where the semiclassical arguments are likely to break down. 

In the case of a complete evaporation a number of exotic physical processes such as 
violations of baryon and lepton number conservation or the ``information paradox''
could occur \cite{wald}. Let us consider a quantum field on the black hole
spacetime whose initial state is described by a pure density matrix $\hat\rho$. 
If we trace over the field modes which are localized inside the event horizon
we are left with an effective mixed state density matrix 
$\hat\rho_{\rm eff}$ for the physics outside the horizon. Of course this does not mean that
a pure state has evolved into a mixed state since the incomplete information 
provided by $\hat\rho_{\rm eff}$ still can be supplemented by the information 
contained in $\hat\rho$ about the degrees of freedom behind the horizon. However, if the 
black hole evaporates completely those parts of the spacetime which formerly where interior
to the horizon disappear entirely, and there are no field degrees of freedom left which could ``know'' 
about the information missing in $\hat\rho_{\rm eff}$. As a consequence, 
the initially pure quantum state $\hat\rho$ seems to have evolved into a genuinely mixed state 
$\hat\rho_{\rm eff}$.

Alternatively one could speculate that the evaporation is incomplete, 
{\it i.e.} that it comes to an end when the Schwarzschild radius is
close to the Planck length where the semiclassical results apply no
longer. In this case the final state of the Hawking evaporation might be some kind of 
``cold'' remnant with a mass close to the Planck mass. 

It is clear that the problem of the final state should be addressed
within a consistent theory of quantum gravity. The standard semiclassical 
derivation of the Hawking temperature quantizes only the matter field
and treats the spacetime metric as a fixed classical background. However, 
investigating black holes with a radius not too far above the Planck 
length we must be prepared that quantum fluctuations of the metric
play an important role. The standard perturbative quantization of Einstein
gravity is of little help here since it leads to a non-renormalizable theory. Also 
the more advanced attempts at formulating a fundamental theory of quantum
gravity (string theory, loop quantum gravity, etc.) do not provide us with a satisfactory
answer yet \cite{rovelli}. As a way out we propose in this paper to use the idea
of the Wilsonian renormalization group \cite{wilson}
in order to study quantum effects in the Schwarzschild spacetime. 

Our basic tool will be a Wilson-type effective action $\Gamma_k[g_{\mu\nu}]$
where $k$ is a scale parameter with the dimension of a mass. In a nutshell, 
$\Gamma_k[g_{\mu\nu}]$ is constructed in such a way that, when evaluated at tree level, 
it correctly describes gravitational phenomena, {\it with all loop effects included}, 
whose typical momenta are of the order of $k$. The basic idea is borrowed from the 
block spin transformations which are used in statistical mechanics in order to 
``coarse-grain'' spin configurations of lattice systems. In its simplest formulation, 
when applied to a continuum field theory \cite{polc,ringwald,abo}, 
we are given a field $\phi(x)$ defined on a Euclidean spacetime with  
metric $g_{\mu\nu}$ and dimension $d$. The averaged or ``blocked'' field $\phi_k(x)$ 
is defined by means of
\be\label{1.1}
\phi_{k}(x) = \int d^d y\sqrt{g}(y) \;\rho_k (x-y) \;\phi(y)
\ee
where $\rho_k(x-y)$ is a smearing function that has support only for $||x-y||<k^{-1}$.
The ``average action'' $\Gamma_k$ governs the dynamics of the coarse-grained or macroscopic
field $\Phi$. It is obtained
from the classical action by integrating  over the microscopic degrees of 
freedom or ``fast variables'':
\be\label{1.2}
\exp(-\Gamma_k[\Phi])=\int D[\phi]\; \delta (\phi_k-\Phi) \exp (-S[\phi])
\ee
The blocked field has a very intuitive physical interpretation: it is  
the field noticed by an observer who uses an experimental apparatus of resolution 
\be\label{1.3}
l \sim k^{-1}
\ee
This observer sees the field evolving according to the effective equation
of motion $\delta \Gamma_k[\Phi]/\delta \Phi (x)=0$. 

For continuum field theories the functional integral (\ref{1.2}) is not easy to deal with, 
and so we shall use an alternative construction which leads to a functional $\Gamma_k$ with 
similar qualitative properties as the one discussed above.  We use the method of the 
``effective average action'' $\Gamma_k$ which has been developed in 
refs.\cite{wet,reuter}. It is defined in a similar way as the ordinary effective action 
$\Gamma$ but it has the additional feature of a built-in infrared cutoff at the scale $k$.
Quantum fluctuations with momenta $p_\mu^2>k^2$ are integrated out in the usual way while
the effect of the large distance fluctuations with $p^2_\mu<k^2$ is not
included in $\Gamma_k$. Hence $\Gamma_k$, regarded as a function of $k$, describes a renormalization
group trajectory in the space of all actions; it connects the classical action 
$S=\Gamma_{k\rightarrow \infty}$ to the ordinary effective action $\Gamma = \Gamma_{k=0}$. 
This trajectory satisfies an exact functional renormalization group (or flow)
equation. If one wants to quantize a fundamental theory with action $S$ one integrates
this equation from the initial point $\Gamma_{ \Lambda}=S$ down to $\Gamma=\Gamma_{ k=0}$.
After appropriate renormalizations one then lets $\Lambda\rightarrow\infty$. 

The flow equation can also be used in order to further evolve (coarse-grain) 
effective field theory actions from one scale $k$ to another. In this case no limit such as
$\Lambda\rightarrow \infty$ above needs to be taken, {\it i.e.} the ultraviolet cutoff
is not removed. Hence the evolution of $\Gamma_k$ from $k_1$ to $k_2$ is always well
defined even if the theory under consideration, when regarded as a fundamental theory, 
is not renormalizable. 

In the following we consider Einstein gravity as an effective field
theory and we identify the standard Einstein-Hilbert action with the average action
$\Gamma_{k_{\rm obs}}$. Here ${k_{\rm obs}}$ is some typical ``observational scale'' 
at which the classical tests of general relativity have confirmed the validity of 
the Einstein-Hilbert action. In order to find an approximate solution to the flow equation we
assume that also for $k>k_{\rm obs}$,  {\it i.e.} at higher momenta, 
$\Gamma_{k}$ is well approximated by an action of 
the Einstein-Hilbert form.  
The two parameters in this action, Newton's constant and the cosmological
constant, will depend on $k$, however, and the flow equation will tell us 
how the running Newton constant $G(k)$ and the running cosmological constant
$\lambda(k)$ depend on the cutoff.  Their experimentally observed values
are $G(k_{\rm obs})=G_0$ and $\lambda(k_{\rm obs})=\lambda_0 \approx 0$.
Since, at least within our approximation, there
is essentially no running of these parameters between $k_{\rm obs}$ (the scale of 
the solar system, say) and cosmological scales ($k \approx 0 $) we may set 
$k_{\rm obs}=0$ and identify the measured parameters with $G(k=0)$ and $\lambda(k=0)$. 

The key idea presented in this paper is to use the running Newton constant $G=G(k)$
in order to ``renormalization group improve''  the Schwarzschild spacetime. This idea
is borrowed from particle physics. There it is a standard device in order to 
add the dominant quantum corrections to the Born approximation of some scattering
cross section, say. Our implementation of this scheme is similar 
to the renormalization  group based
derivation of the Uehling correction to the Coulomb potential in
massless QED \cite{ue}. One starts from the classical potential 
energy $V_{\rm cl}(r) = e^2/4\pi r$ and replaces $e^2$ by the running
gauge coupling in the one-loop approximation: 
\be\label{uel}
e^2(k)=e^2(k_0)[1-b\ln(k/k_0)]^{-1},\;\;\;\;\; b\equiv e^2(k_0)/6\pi^2.
\ee
The crucial step is to identify the 
renormalization point $k$ with the inverse of the 
distance $r$. This is possible because in the massless theory $r$
is the only dimensionful quantity which could define a scale.
The result of this substitution reads
\be\label{rea}
V(r)=-e^2(r_0^{-1})[1+b\ln(r_0/r) + O(e^4)]/4\pi r
\ee
where the IR reference scale $r_0\equiv 1/k_0$ has to be kept finite in the
massless theory. We emphasize that eq.(\ref{rea}) is the correct (one-loop, massless)
Uehling potential which is usually derived by more conventional perturbative methods
\cite{ue}. Obviously the position dependent renormalization group improvement
$e^2\rightarrow e^2(k)$, $k\propto 1/r$ encapsulates the most important effects
which the quantum fluctuations have on the electric field produced by a point
charge. 

In this paper we propose to ``improve'' the Schwarzschild metric by an analogous 
substitution. We replace the Newton constant by its running counterpart
$G(k)$ with an appropriate position-dependent scale $k=k(r)$
where $r$ is the radial Schwarzschild coordinate. At large distances we shall 
have $k(r)\propto 1/r$ as in QED, but since $G$ is dimensionful there will be deviations
at small distances. 

This approach has also been used in ref. \cite{bore} where the impact  of quantized gravity 
on the Cauchy Horizon singularity occurring in a realistic gravitational collapse has been 
studied. In this work a perturbative approximation of the function $G(k)$ has been employed.
In the present paper we use instead an exact, non-perturbative solution to the evolution
equation for $G(k)$ which follows from the ``Einstein-Hilbert truncation''. 

Our main results about the quantum corrected Schwarzschild spacetime are the following. 
For large masses $M$ the quantum effects are essentially negligible. 
Lowering the mass we find that the radius of the event horizon becomes smaller and that
at the same time a second, inner horizon develops out of the  ($r=0$)-singularity 
which is now timelike. When 
$M$ equals a certain critical mass $M_{\rm cr}$ which is of the order of the Planck mass
the two horizons coincide. For $M<M_{\rm cr}$ there is no horizon at all. The causal 
structure of these spacetimes is similar to the classical \RN spacetimes. It turns out
that while the Hawking temperature is proportional to $1/M$ for very heavy black holes
it vanishes as $M$ approaches $M_{\rm cr}$ from above. This leads to a scenario for
the evaporation process where the Hawking radiation is ``switched off'' once the mass
gets close to $M_{\rm cr}$. This picture suggests that the final state of the evaporation
could be a critical (extremal) black hole with $M=M_{\rm cr}$. 

The rest of this paper is organized as follows. 
In section II we derive the running of the Newton constant from the renormalization 
group equation. In section III the correct identification of the position dependent 
cutoff $k=k(r)$  is discussed. 
In  section IV we ``renormalization group improve'' the eternal black hole spacetime
and discuss its properties in detail. 
In section V we provide an effective matter interpretation of this spacetime. In section 
VI the Hawking temperature is derived and our scenario for the evaporation process is 
presented. In Section VII we obtain an expression for the thermodynamic 
entropy of the quantum black hole, while in section
VIII we discuss the fate of the ($r=0$)-singularity.  The conclusions are contained in 
section IX. In the Appendix we discuss some problems related to the statistical mechanical 
entropy of the quantum black hole. 

\renewcommand{\theequation}{2.\arabic{equation}}
\setcounter{equation}{0}
\section{the running newton constant}
In ref. \cite{reuter} the idea of 
the effective average action \cite{wet,corfu} has been 
used in order to formulate the quantization of ($d$-dimensional, Euclidean) 
gravity and the evolution of 
scale-dependent effective gravitational actions $\Gamma_k[g_{\mu\nu}]$ by means of an exact 
renormalization group equation. Furthermore, in order to find approximate 
solutions to this equation,
the renormalization group flow in the infinite dimensional space of all action functionals has been 
projected on the 2-dimensional subspace spanned by the operators $\sqrt{g}$ and $\sqrt{g}R$
(``Einstein-Hilbert truncation''). Using the background gauge formalism with a background metric 
$\bar{g}_{\mu\nu}$, this truncation of the ``theory space'' amounts to considering only actions 
of the form
\be\label{2.1}
\Gamma_k[g,\bar{g}]=(16\pi G(k))^{-1}
\int d^dx\sqrt{g}\{ -R(g)+2\bar{\lambda}(k)\}+S_{\rm gf}[g,\bar{g}]
\ee
where $G(k)$ and $\bar{\lambda}(k)$ denote the 
running Newton constant and cosmological constant, respectively,
and $S_{\rm gf}$ is the classical background gauge fixing term.
For truncations of this type the flow equation reads
\ba\label{2.2}
&&\hskip 2cm \partial_ t \Gamma_k  [g,\bar{g}] = \nonumber\\[2mm]
&&{1\over 2}{\rm Tr}
\Big [( \kappa^{-2} \Gamma^{(2)}_k [g,\bar{g}]+{\cal R}^{\rm grav}_{k}[\bar{g}])^{-1}
\partial_t {\cal R}^{\rm grav}_{k}[\bar{g}]\Big]
-{\rm Tr}\Big[(-{\cal M}[g,\bar{g}]+{\cal R}^{\rm gh}_{k}[\bar{g}])^{-1}
\partial_t {\cal R}^{\rm gh}_k[\bar{g}] \Big ]
\ea
with 
\be\label{2.3}
t \equiv \ln k
\ee
where $\Gamma_k^{(2)}$ stands for the
Hessian of $\Gamma_k$ with respect to 
$g_{\mu\nu}$, and ${\cal M}$ is the Faddeev-Popov 
ghost operator. The operators ${\cal R}_k^{\rm grav}$ 
and ${\cal R}^{\rm gh}_k$ implement the IR cutoff in the graviton
and the ghost sector. They are defined in terms of 
a to some extent arbitrary smooth function ${\cal R}_k(p^2)\propto k^2 R^{(0)}(p^2/k^2)$
by replacing the momentum square $p^2$ with the graviton and ghost kinetic
operator, respectively. Inside loops, they suppress the contribution
of infrared modes with covariant momenta $p<k$. The function $R^{(0)}(z),z\equiv p^2/k^2$,
has to satisfy the conditions $R^{(0)}(0)=1$ and $R^{(0)}(z)\rightarrow 0$ for 
$z \rightarrow \infty$. For explicit computations we use the exponential cutoff
\be\label{2.4}
R^{(0)}(z)=z[\exp(z)-1]^{-1}
\ee

If we insert (\ref{2.1}) into (\ref{2.2}) and project the flow onto the subspace spanned by the 
Einstein-Hilbert truncation we obtain a coupled system of differential equations for the 
dimensionless Newton constant
\be\label{2.5}
g(k)\equiv k^{d-2}G(k)
\ee
and the dimensionless cosmological constant $\lambda(k)\equiv\bar{\lambda}(k)/k^2$:
\be\label{2.6}
\partial_t g = \left[d-2+ \eta_N \right]g
\ee
\be\label{2.7}
\bay\dis
\partial_t\lambda
&=&\dis
-(2-\eta_N)\lambda+\frac{1}{2}g (4\pi)^{1-d/2}\cdot
\\
&&\dis
\quad
\cdot\left[
2d(d+1) \Phi^1_{d/2}(-2\lambda)-8d \Phi^1_{d/2}(0)
-d(d+1)\eta_N\widetilde\Phi^1_{d/2}(-2\lambda)\right]
\eay
\ee
Here
\be\label{2.8}
\eta_N(g,\lambda)= { g B_1(\lambda)\over 1-g B_2(\lambda)}
\ee
is the anomalous dimension of the operator $\sqrt{g} R$, and the functions $B_1(\lambda)$ and
$B_2(\lambda)$ are given by
\be\label{2.9}
\bay\dis
B_1(\lambda)
&\equiv&\dis
\frac{1}{3}(4\pi)^{1-d/2}
\Bigg[
 d(d+1) \Phi^1_{d/2-1}(-2\lambda)
-6d(d-1)\Phi^2_{d/2}(-2\lambda)
\\
&&\dis
\qquad\qquad
\quad
-4d\Phi^1_{d/2-1}(0)-24\Phi^2_{d/2}(0)\Bigg]
\\
B_2(\lambda)
&\equiv&\dis
-\frac{1}{6}(4\pi)^{1-d/2}
\left[
 d(d+1) \widetilde\Phi^1_{d/2-1}(-2\lambda)
-6d(d-1)\widetilde\Phi^2_{d/2}(-2\lambda)
\right]
\eay
\ee
with the cutoff -, {\it i.e.} $R^{(0)}$ - dependent ``threshold'' functions 
$(p=1,2,\cdots)$ 
\be\label{2.10}
\bay\dis
\Phi^p_n(w)
&=&\dis
\frac{1}{\Gamma(n)}\int_0^\infty dz\,
z^{n-1}
\frac{R^{(0)}(z)-z R^{(0)\,\prime}(z)}{[z+R^{(0)}(z)+w]^p}
\\
\widetilde\Phi^p_n(w)
&=&\dis
\frac{1}{\Gamma(n)}\int_0^\infty dz\,
z^{n-1}
\frac{R^{(0)}(z)}{[z+R^{(0)}(z)+w]^p}
\eay
\ee
For further details about the effective average action in gravity and the derivation of the 
above results we refer to \cite{reuter}.

From now on we shall focus on $d=4$. Furthermore, the cosmological constant plays no role within the 
scope of our present investigation. We assume that $\bar{\lambda}\ll k^2$ for all scales of interest 
so that we may approximate $\lambda\approx 0$ in the arguments of $B_1(\lambda)$ and $B_2(\lambda)$.
Thus the evolution is governed by the equation
\be\label{2.11} 
{dg(t)\over dt} = [2+\eta_N]g(t)=\beta(g(t))
\ee
with the anomalous dimension
\be\label{2.12}
\eta_N(g) = {B_1 g\over 1- B_2 g}
\ee
and the beta function
\be\label{2.13}
\beta(g) = 2g\;{1-\omega' \;g\over 1-B_2 \; g }
\ee
The constants $B_1$ and $B_2$ are given by 
\ba\label{2.14}
&& B_1\equiv B_1(0) = -{1\over 3\pi}[24 \Phi^2_2(0)-\Phi^1_1(0)]\\[2mm]
&& B_2\equiv B_2(0) = {1\over 6\pi}[18\tilde{\Phi}^2_2(0)-5\tilde{\Phi}^1_1(0)]
\ea
We also define 
\be\label{2.15}
\omega \equiv -{1\over 2}B_1, \;\;\;\;\;\;\; \omega' \equiv \omega+B_2
\ee 
For the exponential cutoff (\ref{2.4}) we have explicitly
\ba\label{2.16}
&&\Phi^1_1(0)={\pi^2\over 6},\;\;\;\;\;\Phi^2_2(0)=1\\[2mm]
&&\tilde{\Phi}^1_1(0)=1, \;\;\;\;\;\;\; \tilde{\Phi}^2_2(0)={1\over 2}
\ea
and
\be
\omega = {4\over \pi}\Big (1-{\pi^2\over 144} \Big ),~~~~~ B_2 = {2\over 3\pi}
\ee

The evolution equation (\ref{2.11}) displays two fixed points $g_\ast$, $\beta(g_\ast)=0$. There
exists an infrared attractive (gaussian) fixed point at $g_\ast^{\rm IR}=0$ and an ultraviolet attractive
(nongaussian) fixed point at 
\be
g_\ast^{\rm UV} = {1\over \omega'}
\ee
This latter fixed point is a higher dimensional analog of the Weinberg fixed point \cite{we}
known from $(2+\epsilon)$-dimensional gravity. (Within the present framework it has been studied in
\cite{reuter}.)

The UV fixed point separates a weak coupling regime $(g<g_\ast^{\rm UV})$ from a strong coupling regime 
where $g>g_\ast^{\rm UV}$. Since the $\beta$-function is positive for $g\in [0,g_\ast^{\rm UV}]$ and 
negative otherwise, the renormalization group trajectories which result from (\ref{2.11}) with 
(\ref{2.13}) fall into the following three classes:
\begin{description}
\item{({\it i})}~~~Trajectories with $g(k)<0$ for all $k$. They are attracted towards $g_\ast^{\rm IR}$
for $k\rightarrow 0$.
\item{({\it ii})}~~Trajectories with $g(k)>g_{\ast}^{\rm UV}$ for all $k$. They are attracted towards
$g_\ast^{\rm UV}$ for $k\rightarrow\infty$.
\item{({\it iii})} Trajectories with $g(k)\in [0,g_\ast^{\rm UV}]$ for all $k$. They are attracted
towards $g_{\ast}^{\rm IR}=0$ for $k\rightarrow 0$ and towards $g_\ast^{\rm UV}$ for $k\rightarrow\infty$.
\end{description}
Only the trajectories of type $(iii)$ are relevant for us. We shall not allow for a negative Newton
constant , and we also discard solutions of type $(ii)$. They are in the strong coupling region and do not 
connect to a perturbative large distance regime. 
(See ref. \cite{souma} for a numerical investigation of the phase diagram.)

The differential equation (\ref{2.11}) with (\ref{2.13}) can be integrated analytically to yield
\be\label{2.19}
{g\over (1-\omega' g)^{\omega/\omega'}}=
{g(k_0)\over [1-\omega'g(k_0)]^{\omega/\omega'}}\; \Big ({k\over k_0}\Big)^2,
\ee
but this expression cannot be solved for $g=g(k)$ in closed form. However, it is obvious that 
this solution interpolates between the IR behavior $g(k)\propto k^2$ for $k^2\rightarrow 0$ and
$g(k)\rightarrow 1/\omega'$ for $k\rightarrow \infty$.

In order to obtain an approximate analytic expression for the running Newton constant we observe that
the ratio $\omega'/\omega$ is actually very close to unity. Numerically one has 
$\omega\approx 1.2$, $B_2\approx 0.21$, $\omega'\approx 1.4$, $g_\ast^{\rm UV}\approx 0.71$ so that 
$\omega' /\omega \approx 1.18$ is indeed close to 1. Replacing $\omega'/\omega \rightarrow 1$ in 
eq.(\ref{2.19}) yields a rather accurate approximation with the same general features as the exact 
solution. In this case we can easily solve eq.(\ref{2.19}):
\be\label{2.20}
g(k) = {g(k_0)k^2\over \omega g(k_0)\; k^2+[1-\omega g(k_0)]\; k^2_0}
\ee
This function is an {\it exact} solution to the renormalization group equation with the 
approximate anomalous dimension $\eta_N=-2\omega g+O(g^2)$ which is the first term in the 
perturbative expansion of eq.(\ref{2.12}):
\be\label{2.21}
\eta_N = -2\omega g \; \Big [ 1+\sum_{n=1}^\infty (B_2 \; g)^n \Big ].
\ee
Remarkably, for the trajectory (\ref{2.20}) the quantity $B_2~g(k)$ remains negligibly
small for all values of $k$. It assumes its largest value at the UV fixed point where 
$B_2~g_\ast^{\rm UV}=0.15$. Thus equation (\ref{2.20}) provides us with a consistent approximation.
(This can also be checked by comparing to the numerical solution of ref.\cite{souma}.)

In terms of the dimensionful Newton constant $G(k)\equiv g(k)/k^2$ eq.(\ref{2.20}) reads
\be\label{2.22}
G(k) = {G(k_0)\over 1+\omega \;G(k_0)\; [k^2-k_0^2]}
\ee
From now on we shall set $k_0=0$ for the reference scale. At least within the Einstein-Hilbert
truncation, $G(k)$ does not run any more between scales where the Newton constant was determined
experimentally (laboratory scale, scale of the solar system, etc.) and $k\approx 0$ (cosmological
scale). Therefore we can identify $G_0 \equiv G(k_0=0)$ with the experimentally observed value of 
the Newton constant. From
\be\label{2.23}
G(k)={G_0\over 1+\omega \; G_0 \; k^2}
\ee
we see that when we go to higher momentum scales $k$, $G(k)$ decreases monotonically. For small 
$k$ we have
\be
G(k) = G_0-\omega \; G_0^2 \; k^2 +O(k^4)
\ee
while for $k^2\gg G_0^{-1}$ the fixed point behavior sets in and $G(k)$ ``forgets'' its
infrared value:
\be\label{2.25}
G(k)\approx {1\over \omega k^2}
\ee
In ref.\cite{pol}, Polyakov had conjectured an asymptotic running of precisely this form.
\renewcommand{\theequation}{3.\arabic{equation}}
\setcounter{equation}{0}
\section{identification of the infrared cutoff}
In the Introduction we identified the scale $k$ with the inverse distance in order to derive 
the leading QED correction to the Coulomb potential. In this section we discuss how in the case of a 
black hole $k$ can be converted to a position dependent quantity.
We write this position-dependent IR-cutoff in the form
\be\label{3.1}
k(P) = {\xi\over d(P)}
\ee
where $\xi$ is a numerical constant to be fixed later and $d(P)$ is the distance scale
which provides the relevant cutoff for the Newton constant when 
the test particle is located at the point $P$ of the black hole spacetime.

Using Schwarzschild coordinates $(t,r,\theta,\phi)$ and considering spherically 
symmetric spacetimes, the symmetries of the problem imply that $d(P)$ depends on the
$r$-coordinate of $P$ only, $d=d(r)$.

If the test particle is far outside the horizon of the black hole $(r\gg 2 G_0 M)$ where the spacetime
is almost flat we expect that $d(r)$ is approximately equal to $r$. By comparison with the work of
Donoghue \cite{don} we shall see that this is actually the case. As a consequence, 
the function $d$ is normalized such that 
\be\label{3.2}
\lim_{r\rightarrow\infty} {d(r)\over r}=1
\ee
so that the constant $\xi$ fixes the asymptotic behavior
\be\label{3.3}
k(r)\approx {\xi\over r}\;\;\;\;\;{\rm for} \;\;r\rightarrow \infty
\ee

Contrary to the situation in QED on flat spacetime, eq.(\ref{3.3}) is not a satisfactory
identification of $k=k(P)$ for arbitrary points $P$. The 
reason is that $d(P)$ should have a coordinate independent
meaning, while $r$ is simply one of the local Schwarzschild coordinates. As a way out, we define 
$d(P)$ to be the proper distance (with respect to the classical Schwarzschild metric) from 
the point $P$ to the center of the black hole along some curve $\cal C$:
\be\label{3.4}
d(P) = \int_{\cal C} \sqrt{|ds^2|}
\ee
There is still some ambiguity as for the correct identification of the spacetime curve $\cal C$.
However, at least in the spherically symmetric case, it turns out that all 
physically plausible candidates lead to cutoffs with the same qualitative features.   

We parametrize ${\cal C}$ as $x^\mu(\lambda)$ where $x^\mu=(t,r,\theta,\phi)$ are the Schwarzschild
coordinates and $\lambda$ is a (not necessarily affine) parameter along the curve. To start with, let us 
consider the curve ${\cal C}\equiv {\cal C}_{(1)}$ defined by 
$t(\lambda)=t_0$, $r(\lambda)=\lambda$, $\theta(\lambda)=\theta_0$, $\phi(\lambda)=\phi_0$ with 
$\lambda\in [0,r(P)]$ where $r(P)$ is the $r$ coordinate of $P$. This is, even beyond the horizon, 
a straight ``radial'' line from the origin to $P$, at fixed values of $t,\theta$ and $\phi$. If we restrict
$\lambda$ to the interval $[r(P_0),r(P)]$ with $P_0$ and $P$ both outside the horizon where 
$ds^2>0$ then $\int\sqrt{ds^2}$ is the ordinary spatial proper distance between the points $P_0$ and 
$P$. The definition (\ref{3.4}) involves the modulus of $ds^2$ and it generalizes this ``distance''
to the case that at least one of the two points lies within the horizon where $r$ is timelike.  
(See also \cite{MTW} for a discussion of this ``distance''.) The explicit result reads for $r<2 G_0 M$
\be\label{3.5}
d_{(1)}(r) = 2 G_0M \; {\rm arctan}\sqrt{r\over 2G_0M-r}-\sqrt{r(2G_0M-r)}
\ee
and for $r>2G_0M$:
\be\label{3.6}
d_{(1)}(r) = \pi G_0M+2G_0M \;\ln\left ( \sqrt{r\over 2G_0M}+\sqrt{{r\over 2G_0M}-1}\right )
+\sqrt{r(r-2G_0M)}.
\ee
Note that $d_{(1)}(r)$ is continuous at the horizon. 
Eq.(\ref{3.6}) shows that indeed $d_1(r)=r+O(\ln r)$ for
$r\rightarrow\infty$. From (\ref{3.5}) we obtain for $r\rightarrow 0$ 
\be\label{3.7}
d_{(1)}(r)={2\over 3}{1\over \sqrt{2G_0M}}\;r^{3/2}+O(r^{5/2})
\ee
which leads to the cutoff 
\be\label{3.8}
k_{(1)}(r)={3\over2}\;\xi\;\sqrt{2G_0M}\;\left ({1\over r}\right)^{3/2}\;\;\;{\rm for}\; r\rightarrow 0.
\ee
This $r^{-3/2}$-behavior has to be contrasted with the $r^{-1}$-dependence of the ``naive'' cutoff
$k=\xi/r$. 

Another plausible spacetime curve ${\cal C}$ is the worldline of an observer who falls into 
the black hole. We define ${\cal C}\equiv {\cal C}_{(2)}$ to be the radial timelike geodesic of the 
Schwarzschild metric with vanishing velocity at infinity. For this geodesic, the observer's
radial coordinate $r$ and proper time $\tau$ are related by \cite{MTW} 
\be\label{3.9}
\tau-\tau_0 = {2\over 3}{1\over \sqrt{2G_0M}}\;(r_0^{3/2}-r^{3/2})
\ee
where the constant of integration is chosen such that $r({\tau_0})=r_0$. Eq.(\ref{3.9}) is valid 
both outside and inside the horizon. Setting $r_0=0=\tau_0$, we see that when the observer has arrived at 
$r=r(P)$, the remaining proper time it takes him or her to reach the singularity is given by
\be\label{3.10}
|\tau(P)|={2\over 3}{1\over \sqrt{2G_0M}}\;r(P)^{3/2}
\ee
From the point of view of this observer it is meaningless to consider times larger than 
$|\tau(P)|$ and, as a consequence, frequencies smaller than $|\tau(P)|^{-1}$. This motivates the
identification $d_{(2)}(P)=|\tau(P)|$, {\it i.e.}
\be\label{3.11}
d_{(2)}(r)={2\over 3}{1\over \sqrt{2G_0M}}\;r^{3/2}
\ee
which leads to 
\be\label{3.12}
k_{(2)}(r)={3\over 2}\;\xi\; \sqrt{2G_0M}\;\left ({1\over r}\right)^{3/2}
\ee
Eqs.(\ref{3.11}) and (\ref{3.12}) are exact for all values of $r$. It is remarkable that 
$d_{(2)}(r)$ coincides precisely with the approximation for $d_{(1)}(r)$, eq.(\ref{3.7}),
which is valid for small values of $r$. This supports our assumption that close to the singularity 
$(r\rightarrow 0)$ the correct cutoff behaves as $k(r)\propto 1/r^{3/2}$.

For large distances, the curves ${\cal C}_{(1)}$ and ${\cal C}_{(2)}$ lead to 
different $r$-dependencies 
of the cutoff: $k_{(1)}\propto 1/r$, $k_{(2)}\propto 1/ r^{3/2}$. Quite generally, if a system
possesses more than one typical momentum scale, $k_{(1)}$, $k_{(2)}$, $k_{(3)}, \cdots$ which can cut off
the running of some coupling constant, it is the largest one among those scales which provides the 
actual cutoff: $k={\rm Max}\{$  $k_{(1)}$, $k_{(2)}$, $k_{(3)}, \cdots  \}$. In the case at hand we have
$k_{(1)}\gg k_{(2)}$ for $r\rightarrow \infty$ so that we must set $k=k_{(1)}(r)\propto 1/r$ 
for large values of $r$. 

The only properties of the function $k(r)$ which we shall use in the following is that it varies
as $k(r)\propto 1/r$ for $r\rightarrow \infty$ and as $k(r)\propto 1/r^{3/2}$ for $r\rightarrow 0$.
This behavior can be further confirmed by investigating different choices of ${\cal C}$.
For instance, a radial timelike geodesic with vanishing velocity at some finite distance from the black hole
or a geodesic with non-vanishing velocity at infinity, for small values of $r$, again reproduces 
(\ref{3.7}).

While we used Schwarzschild coordinates in the above discussion we emphasize that the same results can
also be obtained using coordinate systems (such as the Eddington-Finkelstein coordinates) which do not
become singular at the horizon. 

It turns out that the qualitative features of the quantum corrected black hole spacetimes which we are 
going to construct
in the following are rather insensitive to the precise manner in which $k(r)$ interpolates between the
$1/r^{3/2}$ and the $1/r$ behavior. Moreover, most of the general features (horizon structure, etc.)
are even independent of the precise form of $k(r)$ for $r\rightarrow 0$. Using $k(r)\propto 1/r^\nu$
with $\nu$ not necessarily equal to $3/2$ leads to essentially the same picture. The only issue where 
the value of $\nu$ is of crucial importance is the fate of the singularity at $r=0$ when quantum
effects are switched on.

In concrete calculations we shall use the interpolating function
\be\label{3.13}
d(r)=\left ( {r^3\over r+\gamma \; G_0\; M} \right )^{1\over 2}
\ee
with $d(r)=r[1+O(1/r)]$ and $d(r)=r^{3/2}/\sqrt{\gamma G_0 M}+O(r^{5/2})$ for large and small 
$r$'s, respectively. From ${\cal C}_{(1)}$ and ${\cal C}_{(2)}$ we had obtained 
\be\label{3.14}
\gamma = {9\over 2}
\ee
but we shall treat $\gamma$ as a free parameter. Most of our results turn out to be very robust:
qualitatively they are the same for all $\gamma >0$. By setting $\gamma=0$, the ansatz (\ref{3.13})
also allows us to return to the ``naive'' cutoff $k\propto 1/r$, {\it i.e.} to $\nu=1$.
Except for questions related to the singularity at $r=0$, even $\gamma=0$ will lead to essentially 
the same qualitative properties of the improved black hole spacetime. 

Upon inserting (\ref{3.1}) into the running Newton constant (\ref{2.23}) we obtain the following 
position-dependent Newton constant $G(r)\equiv G(k(r))$:
\be\label{3.15}
G(r)={G_0 \; d(r)^2\over d(r)^2+\tilde{\omega}\; G_0}
\ee
where
\be\label{3.16}
\tilde{\omega}\equiv\omega\xi^2
\ee
For the ansatz (\ref{3.13}) this yields
\be\label{3.17}
G(r)={G_0 \; r^3\over r^3 +\tilde{\omega}\;G_0\; [r+\gamma G_0 M]}
\ee
At large distances, the leading correction to Newton's constant is given by 
\be\label{3.18}
G(r)=G_0-\tilde{\omega}\;{G_0^2\over r^2}+O({1\over r^3}).
\ee
For small distances $r\rightarrow 0$, it vanishes very rapidly:
\be\label{3.19}
G(r)={r^3\over \gamma \tilde{\omega} G_0 M}+O(r^4)
\ee

The asymptotic behavior (\ref{3.18}) can be used in order to fix the numerical
value of $\tilde{\omega}$. The idea is to renormalization group improve the
classical Newton potential $V(r)=-G_0 m_1 m_2 /r$ of two masses $m_1$ and $m_2$ at 
distance $r$ by replacing the constant $G_0$ with $G(r)$. Within the approximation 
(\ref{3.18}) we obtain
\be\label{3.20}
V_{\rm imp}(r)=-G_0{m_1 m_2\over r}\left [1-\tilde{\omega} \; {G_0\hbar\over r^2 c^3}+\cdots\right ]
\ee
where we have reinstated factors of $\hbar$ and $c$ for a moment. We observe that our renormalization
group approach predicts a $1/r^3$-correction to the $1/r$-potential. However, the value of the 
coefficient $\tilde{\omega}=\omega\xi^2$ cannot be obtained by renormalization group arguments alone:
the factor $\omega$ is a non-universal coefficient of the $\beta$-function, {\it i.e.} it depends
on the shape of the function $R^{(0)}$, and also $\xi$ is unknown as long as one does not explicitly
identify the specific cutoff for a concrete process.

On the other hand, it was pointed out by Donoghue \cite{don} that the standard perturbative 
quantization of Einstein gravity leads to a well-defined, finite prediction for the leading
large distance correction to Newton's potential. His result reads
\be\label{3.21}
V(r)=-G_0{m_1 m_2\over r}\;\left  [ 1-{G_0(m_1+m_2)\over 2 c^2 r}-\hat{\omega}\;
{G_0\hbar\over r^2 c^3}+\cdots\right ]
\ee
where \cite{hamber} $\hat{\omega}=118/15\pi$. The correction proportional to 
$(m_1+m_2)/r$ is a purely kinematic effect of classical general relativity, while
the quantum correction $\propto \hbar$ has precisely the structure we have predicted on the
basis of the renormalization group. Comparing (\ref{3.20}) to (\ref{3.21}) allows us to determine the 
coefficient $\tilde{\omega}$ by identifying 
\be\label{3.22}
\tilde{\omega}=\hat{\omega}\equiv{118\over 15 \pi}.
\ee
Contrary to the factors $\omega$ and $\xi^2$, their product $\tilde{\omega}=\omega\xi^2$ has a
uniquely determined, measurable value. 

A priori the renormalization group analysis yields $G$ as a function
of $k$ rather than $r$, and the function ${\cal R}_k$ serves as a mathematical model of an arbitrary, 
yet unspecified physical mechanism which cuts off the running of $G$. In the case at hand, this mechanism
is the finite distance between the test particle and the black hole; it led to the ansatz $k=\xi/d(r)$.
In general the information about the actual physical cutoff mechanism enters at two points:
\begin{description}
\item{a)}\hspace{2mm} The function ${\cal R}_k$ should be chosen so as to model
the actual physics as correctly as possible.
\item{b)}\hspace{2mm} Both the physical cutoff mechanism and the choice for ${\cal R}_k$ determine
the relation between $k$ and other variables, adapted to the concrete problem, which 
can parametrize the running of $G$ ($r$, in our case).
\end{description}
This means that, within our approximation, the ${\cal R}_k$-dependence of the 
correct identification $k=k(r)$ should precisely
compensate for the ${\cal R}_k$-dependence of $G(k)$. We have seen that this is indeed what happens:
$\omega$ and $\xi$ appear only in the combination $\tilde{\omega}=\omega\xi^2$.
The ${\cal R}_k$-dependencies of $\omega$ and $\xi^2$ cancel in this product, and its unambiguous 
numerical value can be read off from the known asymptotic form of $V_{\rm imp}(r)$.
\renewcommand{\theequation}{4.\arabic{equation}}
\setcounter{equation}{0}
\section{improving the eternal black hole spacetime}
\subsection{The improved metric}
We consider  spherically symmetric, Lorentzian metrics of the form
\be\label{4.1}
ds^2=-f(r)\; dt^2+f(r)^{-1}\; dr^2+r^2d\Omega^2
\ee
where $d\Omega^2\equiv d\theta^2+\sin^2\theta d\phi^2$ is the line element on the unit two-sphere
and $f(r)$ is an arbitrary ``lapse function''.
The most important example of a metric belonging to this class is the Schwarzschild metric with
\be\label{4.2}
f(r) = f_{\rm class}(r) \equiv 1-{2 G_0 M\over r}
\ee
While the Schwarzschild spacetime is a solution of the vacuum Einstein equation $R_{\mu\nu}=0$,
we are not going to constrain $f(r)$ by any field equation in the following.

In classical general relativity the metric (\ref{4.1}) with (\ref{4.2}) is interpreted as a 
property of a black hole (or the exterior of a star) {\it per se}, i.e. the metric is given a meaning 
even in absence of a test particle which probes it. Within our approach, we regard $f_{\rm class}$
as a manner of encoding the classical dynamics of a test particle in the vicinity of some ``central
body'' of mass $M$. 
Because of the actual presence of the test particle, the system defines a physically 
relevant distance scale $d(r)$ which enters into the cutoff for the running of $G$.
It is our main assumption that, also beyond the Newtonian limit, the leading quantum 
gravity effects in this system consist of a position dependent renormalization of the 
Newton constant in (\ref{4.2}). More precisely, we assume that the quantum corrected 
geometry can be approximated by (\ref{4.1}) with 
\be\label{4.3}
f(r)=1-{2 G(r) M\over r}
\ee
where $G(r)$ is given by (\ref{3.17}):
\be\label{4.4}
f(r)=1-{2 G_0 M \; r^2 \over r^3 +\tilde{\omega}G_0 \; [r+\gamma G_0 M]}
\ee

Let us now analyse the properties of the renormalization group improved spacetime defined by
Eq.(\ref{4.4}). First of all, for $r\rightarrow \infty$ we have 
\be\label{4.4a}
f(r)=1-{2 G_0 M\over r}\left ( 1-\tilde{\omega} {G_0\over r^2}\right )
+ O\left ({1\over r^4}\right )
\ee
For large distances, {\it i.e.} at order $1/r$, we recover the classical Schwarzschild 
spacetime. The leading quantum correction appears at order $1/r^3$; since in the Newtonian
approximation the potential is given by $[f(r)-1]/2$, this correction is equivalent to the 
improved potential (\ref{3.20}) which was independently confirmed by Donoghue's 
result (\ref{3.21}). As we discussed in section III already, matching the two results 
unambiguously fixes the constant $\tilde\omega$ to be $\tilde\omega=\hat\omega=118/15\pi$. 
Thus our improved lapse function (\ref{4.4}) does not contain any free parameter. 
(Recall that the analysis of section III fixes $\gamma$ to be  $\gamma = 9/2$. However, to be as general 
as possible, we shall allow for an arbitrary $\gamma\geq 0$ in  most of the calculations.)
\subsection{The horizons}
Next we determine the structure of the horizons of the improved spacetime. To this end 
we look for zeros of the function $f(r)$,  eq. (\ref{4.4}),  which is conveniently rewritten as
\be\label{4.5}
f(r) = {B(x)\over B(x) + 2x^2}\Big |_{x=r/G_0 M}
\ee
with the polynomial $B$ given by 
\be\label{4.6}
B(x)\equiv B_{\gamma,\Omega}(x)=x^3-2x^2+\Omega \; x +\gamma \; \Omega
\ee
where 
\be\label{4.7}
\Omega\equiv {\tilde{\omega}\over G_0 M^2}
\ee

The parameter $\Omega$ is a measure for the impact the quantum gravity effects have 
on the metric. Reinstating factors of $\hbar$ for a moment we have $\tilde\omega\propto \hbar$ 
and $\Omega\propto\hbar$. The classical limit is recovered by setting $\Omega=0$. We see immediately
that very heavy black holes $(M\rightarrow\infty)$ essentially behave classically. 
In fact, defining the Planck mass by $m_{\rm Pl}\equiv G_0^{-1/2}$ we have 
\be\label{4.8}
\Omega=\tilde{\omega} \; {m_{\rm Pl}^2\over M^2}
\ee
which shows that an $\Omega$ of order unity requires $M$ 
to be not much heavier than $m_{\rm Pl}$.
 
For $x>0$ the numerator and the denominator of the RHS of eq.(\ref{4.5}) have no common zeros;
hence $r_0$ is a zero of $f(r)$ if $x_0=r_0/G_0 M$ is a zero of $B(x)$. In the classical case
$(\Omega=0)$ we have $B_{\gamma,0}(x)=x^2(x-2)$ with its only nontrivial zero $x_0=2$ 
corresponding to the familiar Schwarzschild horizon at $r_0=2 G_0M$.

In the quantum case $(\Omega >0)$, $B_{\gamma, \Omega}(x)$ is a generic cubic polynomial which 
has either one or three simple zeros\footnote{Here double and triple zeros are counted as 
two or three simple zeros, respectively.} on the real axis. Since $r\equiv x G_0 M$ must be 
positive, only zeros on the positive real $x$-axis $\Re^{+}$ can correspond to a horizon.
It is easy to see that for any value of $\Omega$ and $\gamma$, $B(x)$ always has precisely 
one zero on the {\it negative} real axis: first we observe that $B(0)=\gamma \Omega>0$ and 
$B(-\infty)=-\infty<0$ which implies that $B(x)$ has at least one zero on the negative real axis.
Furthermore, the derivative $B'(x)=3x^2-4x+\Omega$ is positive for $x<0$, {\it i.e.} $B$ is 
monotonically increasing for $x<0$. As a consequence, $B$ has precisely one zero on the
negative real axis. Hence $B$ has either two simple zeros or no zeros at all on the positive real axis
$\Re^+$, whereby the two simple zeros might degenerate to form a single double zero.

\begin{figure}
\hbox to\hsize{\hss\epsfxsize=11cm\epsfbox{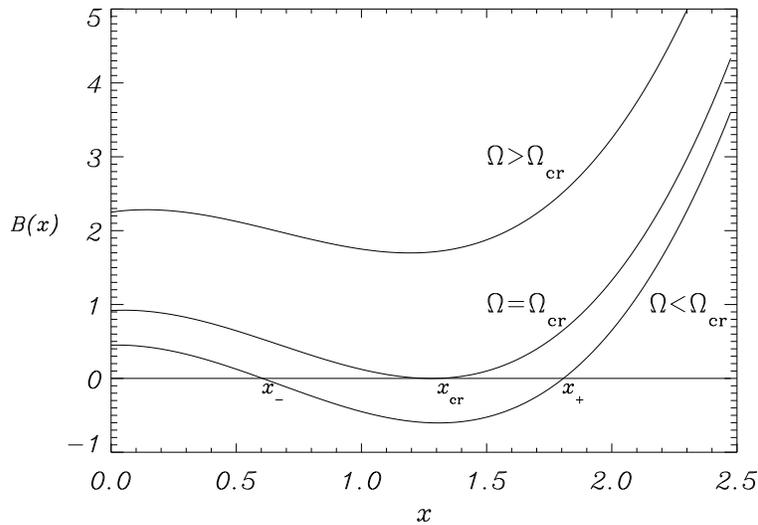}\hss}
\caption{ The function $B_{\gamma, \Omega}(x)$ with $\gamma=9/2$ 
for different values of $\Omega$. The regime $\Omega < \Omega_{\rm cr}$ 
($\Omega > \Omega_{\rm cr}$) corresponds to very heavy (light) black holes.\label{fig1}}
\end{figure}

The three cases are distinguished by the value of the discriminant 
\be\label{4.9}
{\cal D}_\gamma (\Omega) = (3\Omega-4)^3+\left (9\Omega+{27\over 2}\; \gamma \; \Omega-8\right )^2
\ee
For ${\cal D}_\gamma(\Omega)<0$ there are two simple zeros on $\Re^+$, for 
${\cal D}_\gamma(\Omega)=0$ we have a double zero, and for ${\cal D}_\gamma(\Omega)>0$
there exists no zero on $\Re^+$. The discriminant can be factorized as 
\be\label{4.10}
{\cal D}_\gamma (\Omega) = 27\; \Omega\; [\Omega-\Omega_1(\gamma)] \; 
[\Omega-\Omega_{\rm cr}(\gamma)]
\ee
with 
\be\label{4.11}
\Omega_{\rm cr}(\gamma)={1\over 8}(9\gamma+2)\; \sqrt{\gamma+2}\; \sqrt{9\gamma+2}
-{27\over 8 }\; \gamma^2-{9\over 2}\; \gamma+{1\over 2}
\ee
The function $\Omega_1(\gamma)$ is not important except that it is negative for all 
$\gamma>0$. As a consequence, the sign of ${\cal D}_{\gamma}(\Omega)$ depends only on
whether $\Omega$ is smaller or larger than the critical value $\Omega_{\rm cr}$:
For $\Omega<\Omega_{\rm cr}(\gamma)$ the polynomial $B_{\gamma,\Omega}$ has two 
simple zeros $x_{+}$ and $x_{-}$ on $\Re^+$ $(x_{+}>x_{-}>0)$, 
for $\Omega>\Omega_{\rm cr}(\gamma)$ it has no zero on $\Re^+$, and for 
$\Omega=\Omega_{\rm cr}(\gamma)$ the two simple zeros merge into a single double 
zero at $x_{+}=x_{-}\equiv x_{\rm cr}$. This situation is illustrated in Fig.(\ref{fig1}).

By virtue of eq.(\ref{4.7}), the critical value for $\Omega$ defines a critical 
value for the mass of the black hole: 
\be\label{4.12}
M_{\rm cr}(\gamma) = \left [ {\tilde{\omega}\over \Omega_{\rm cr}(\gamma)G_0}
\right ]^{1/2}
\ee
For the preferred value $\gamma = 9/2$ we have 
\be\label{4.13}
\Omega_{\rm cr}({9/ 2})={1\over 32}[85\sqrt{85}\sqrt{13}-2819]\approx 0.20
\ee
while for $\gamma = 0$ (``naive'' cutoff $k\propto 1/r$ )
\be\label{4.13a}
\Omega_{\rm cr}(0) = 1
\ee
In any case $M_{\rm cr}$ is a number of order unity times $m_{\rm Pl}$.

The zeros $x_{\pm}$ or $x_{\rm cr}$ of $B(x)$ are equivalent to zeros of $f(r)$ located at 
\be\label{4.14}
r_{\pm}=x_{\pm} G_0 M, \;\;\;\;\;\;\;\;\; r_{\rm cr}=x_{\rm cr} G_0 M_{\rm cr}
\ee
They correspond to horizons of the quantum corrected black hole spacetime. For heavy black holes
$(M>M_{\rm cr}, \Omega<\Omega_{\rm cr})$ we have an outer horizon at $r_{+}$
and an inner horizon at $r_{-}$. 
The function $f(r)$ is positive, {\it i.e.} the vector field ${\partial / \partial t}$
is timelike outside the outer (event) horizon $(r>r_{+})$ and inside the 
inner horizon $(r<r_{-})$; in the region between the horizons $(r_{-}<r<r_{+})$
we have $f(r)<0$ so that $\partial \over \partial t$ is spacelike. 
\begin{figure}
\hbox to\hsize{\hss\epsfxsize=11cm\epsfbox{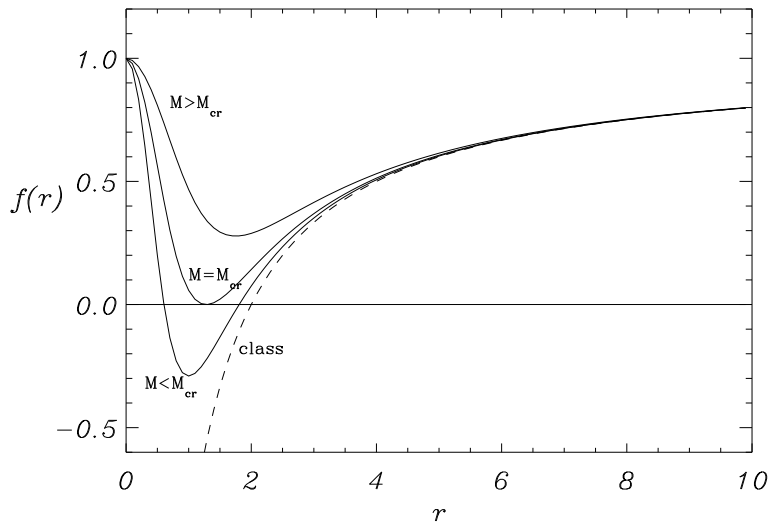}\hss}
\caption{
The lapse function $f(r)$ for various mass values. The dashed line
shows $f_{\rm class}(r)$ of the
classical Schwarzschild metric.
\label{fig2}}
\end{figure}
For $M\gg M_{\rm cr}$ the outer horizon coincides essentially with the classical 
Schwarzschild horizon $(r_{+}\approx 2G_0 M)$ while $r_{-}$ is very close zero. 
When we decrease $M$ and approach $M_{\rm cr}$ from above, the outer horizon shrinks and the inner 
horizon expands. Finally, for $M=M_{\rm cr}$, the two horizons 
coalesce at $r_{+}=r_{-}\equiv r_{\rm cr}$ which corresponds to a double zero of $f$. 
For very light black holes with $M<M_{\rm cr}$ the spacetime has no horizon at all. 

\begin{figure}
\hbox to\hsize{\hss\epsfxsize=11cm\epsfbox{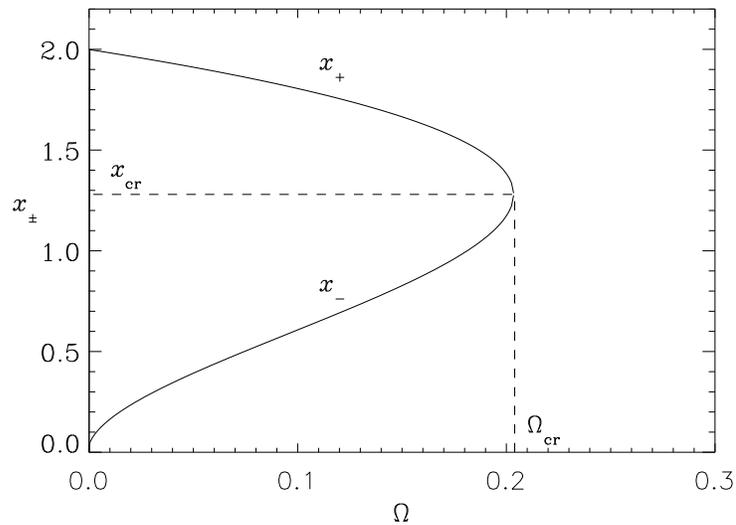}\hss}
\caption{
The zeros $x_{+}$ and $x_{-}$ for $\gamma = 9/2$ as a function of
$\Omega$. Lowering $M$ from infinity to $M_{\rm cr}$, $\Omega$ increases from zero to 
$\Omega_{\rm cr}$, $x_{-}$ increases from zero to $x_{\rm cr}$, and
$x_{+}$ decreases from 2 towards $x_{\rm cr}$. The outer horizon shrinks and the inner horizon
expands until they meet at $r_{\rm cr}$.  
\label{fig3}}
\end{figure}

In Fig.(\ref{fig2}) we plot $f(r)$ for various masses $M$. The values of $x_{+}$ and $x_{-}$ could be 
written down in closed form as a function of $\Omega$ and $\gamma$, but the formulas are not very 
illuminating. Instead, in Fig.(\ref{fig3}), we represent them graphically. 
\subsection{The critical (extremal) black hole and the \RN analogy}
Let us look in more detail at the ``critical'' black hole with $M=M_{\rm cr}$. We know that 
for $\Omega=\Omega_{\rm cr}(\gamma)$ the polynomial $B_{\gamma,\Omega}(x)$ has a double
zero at some $x_{\rm cr}\equiv x_{\rm cr}(\gamma)>0$. Upon inserting eq.(\ref{4.11}) into
$B_{\gamma,\Omega_{\rm cr}}(x)$ and factorizing the resulting expression with respect to $x$ one
finds the following  explicit result:
\be\label{4.14a}
x_{\rm cr}(\gamma)={1\over 4}\sqrt{\gamma+2}\sqrt{9\gamma+2}-{3\over 4}\gamma
+{1\over2}
\ee
In particular, 
\ba\label{4.14b}
&&x_{\rm cr}(0)\;\;\;=\;1\nonumber\\[2mm]
&&x_{\rm cr}(9/2)={1\over 8}\;[\sqrt{13}\sqrt{85}-23]\approx 1.28\nonumber
\ea
Using (\ref{4.14}) the critical radius reads 
\be\label{4.14c}
r_{\rm cr}(\gamma)=x_{\rm cr}(\gamma)\;\left ( 
{\tilde{\omega}G_0\over\Omega_{\rm cr}(\gamma)} \right )^{1/2}
\ee
with $x_{\rm cr}$ and $\Omega_{\rm cr}$ given by (\ref{4.14a}) and ({\ref{4.11}), respectively.

Some of the qualitative features of the quantum black hole are remarkably similar to those of a
\RN spacetime (black hole with charge $e$). Its lapse function reads
(in appropriate units)
\be\label{4.15}
f_{\rm RN}(r) = 1-{2 G_0 M\over r}+{G_0 e^2\over r^2}
\ee
In analogy with (\ref{4.7}) we introduce the parameter 
\be\label{4.16}
{\Omega_{\rm RN}}\equiv {e^2\over G_0 M^2}
\ee
The \RN spacetime has no horizon for $\Omega_{\rm RN}>1$, two horizons
with 
\be\label{4.17}
x_{\pm}^{\rm RN}=r_{\pm}^{\rm RN}/G_0 M = 1\pm\sqrt{1-\Omega_{\rm RN}}
\ee
if $\Omega_{\rm RN}<1$, and a single degenerate horizon at 
\be\label{4.18}
r_{\rm cr}^{\rm RN}=G_0 M
\ee
if $\Omega_{RN}$ equals its critical value $\Omega_{RN}=1$. We observe that, in a sense, 
the renormalization group improved Schwarzschild spacetime is similar to a
\RN black hole whose charge is given by $e=\tilde{\omega}^{1/2}$.
In particular, the ``critical'' quantum black hole with $M=M_{\rm cr}$ corresponds to the
extremal charged black hole $(\Omega_{RN}=1)$. 

Let us look more closely at the near-horizon geometry of the critical quantum 
black hole. If we expand about $r=r_{\rm cr}$ and introduce the new coordinate
$\bar{r}\equiv r-r_{\rm cr}$ we have at leading order
\be\label{rb1}
ds^2=-\left ( {\bar{r}\over G_0 M_{\rm AdS}}\right )^2 \; dt^2
+\left ( {G_0 M_{\rm AdS}\over \bar{r}}\right )^2 \; d\bar{r}^2
+r^2_{\rm cr} \; d\Omega^2
\ee
where the mass parameter $M_{\rm AdS}$ is defined by 
\be
(G_0 M_{\rm AdS})^{-2}={1\over 2} f''(r_{\rm cr}(\gamma))\Big |_{\Omega=\Omega_{\rm cr}(\gamma)}
\ee\label{rb2}
The metric (\ref{rb1}) is the Robinson-Bertotti metric for the product of a two-dimensional 
anti-de Sitter space $AdS_2$ with a two-sphere, $AdS_2\times S^2$. The parameter $M_{\rm AdS}$
determines the curvature of $AdS_2$. Using (\ref{4.4}), (\ref{4.11}) and
(\ref{4.14a}) it is obtained in the form
\be\label{rb3}
M_{\rm AdS}(\gamma) = \sqrt{2/b(\gamma)} \;M_{\rm cr}
\ee
where $b(\gamma)$ is a complicated function which we shall not write down here. In particular,
\ba\label{rb4}
&&b(\gamma=0)\;\;\;=1\\[2mm]
&&b(\gamma = 9/2) = 1.123\cdots
\ea
If we put $M_{\rm AdS} = M$ and $r_{\rm cr}= G_0 M$ the metric (\ref{rb1}) describes also the near-horizon
geometry of an extremal \RN black hole of mass $M$. However, in our case the
relative magnitude of the $AdS_2$ - and the $S^2$ - curvature is different. For 
$\gamma=0$, say, the $S^2$-curvature is given by the radius $r_{\rm cr}=G_0 M_{\rm cr}$
as above, but the $AdS_2$-curvature is determined by $M_{\rm AdS}=\sqrt{2} M_{\rm cr}$. 
\subsection{Large mass expansion of $r_{\pm}$}
It is instructive to look at the location of the horizons in the limit of very heavy black holes.
Since $\Omega\propto 1/M^2$, the large-mass expansion in $1/M$ corresponds to an expansion in powers
of $\Omega^{1/2}$. Let us start by looking at the leading quantum correction of the outer horizon.
Classically we have $r_{+}=2G_0M$ or $x_{+}=2$. By inserting an ansatz of the form
$x_{+}=2+c_1\Omega+c_2\Omega^2+\cdots$ into $B(x_{+})=0$ and combining equal powers of $\Omega$ 
we can easily determine the coefficients $c_j$. In leading order one finds 
$x_{+}=2-{1\over 4}(2+\gamma)\Omega+O(\Omega^2)$ and
\be\label{4.19}
r_{+}=2 G_0 M-{(2+\gamma)\tilde{\omega}\over4 M}+O({1\over M^3})
\ee
We see that the quantum corrected $r_{+}$ is indeed smaller than its classical 
value. The leading correction is proportional to $1/M$ and it is independent 
of the value of Newton's constant. The prefactor of the $1/M$-term is uniquely determined: 
the arguments of section III yield $\gamma = {9/2}$ and $\tilde\omega$ is fixed by the 
matching condition (\ref{3.22}). We believe that (\ref{4.19}) is a particularly accurate 
prediction of our approach.

Let us now look at 
$r_{-}$ for $M\rightarrow\infty$. Classically, for $\Omega=0$, we have $B(x)=x^2(x-2)$.
When we switch on $\Omega$, the double zero at $x=0$ develops into 2 simple zeros, one on
the negative and the other on the positive real axis. The latter is the 
(approximate) $x_{-}$we are looking for. As long as $\Omega\ll 1$ we have $x_{-}\ll 1$
and therefore we may neglect the cubic term in $B(x_{-})=0$ relative to the quadratic one.
The resulting equation is easily solved:
\be\label{4.20}
x_{-}={1\over 4}\; \sqrt{\Omega}\; [\sqrt{\Omega}+\sqrt{8\gamma+\Omega}]
\ee
The asymptotics of this result depends on whether $\gamma>0$ or $\gamma=0$. For
$\gamma > 0$ we have 
\be\label{4.21}
x_{-}={1\over 2}\sqrt{2\gamma \Omega}+O(\Omega)=
\sqrt{{\gamma\tilde{\omega}\over 2 G_0}}\; {1\over M}+O({1\over M^2})
\ee
Obviously $x_{-}$ vanishes for $M\rightarrow \infty$ but because of its $1/M$
behavior the actual radius $r_{-}=x_{-} G_0 M$ approaches a universal, nonzero limit:
\be\label{4.22}
r_{-}={1\over 2}\sqrt{2\gamma\tilde{\omega} G_0}+O({1\over M})
\ee
Thus the inner horizon does not disappear even for infinitely massive black holes.
The situation would be different for $\gamma=0$. There $x_{-}={1\over 2}\Omega+O(\Omega^{3/2})$
and 
\be\label{4.23}
r_{-}={\tilde{\omega}\over 2M}+O({1\over M^2})\;\;\;\;\;(\gamma=0)
\ee
which vanishes for $M\rightarrow \infty$.
\subsection{The de Sitter core}
We expect the improved $f(r)$ to be reliable as long as $r$ is not too close to $r=0$ 
where the renormalization effects become strong and the quantum corrected geometry
differs significantly from the classical one. Therefore eqs.(\ref{4.22}) and 
(\ref{4.23}) should be taken with a grain of salt, of course. However, if one takes
eq.(\ref{4.4}) at face value even for $r\rightarrow 0$, the horizons 
(\ref{4.22}), (\ref{4.23}) acquire a very intriguing interpretation. 

Expanding $f(r)$ about $r=0$ one finds for $\gamma>0$ 
\be\label{4.24}
f(r)=1-2(\gamma\tilde{\omega}G_0)^{-1}\; r^2 +O(r^3)
\ee
Recalling that (\ref{4.1}) with $f_{\rm dS}(r)=1-\Lambda r^2/3$ is the metric of de Sitter 
space we see that, at very small distances, the quantum corrected Schwarzschild spacetime looks like
a de Sitter space with an effective  cosmological constant
\be\label{4.25}
\Lambda_{\rm eff}=6\; (\gamma\tilde{\omega} G_0)^{-1}
\ee
(For $\gamma = 9/2$, $\Lambda_{\rm eff}\approx 0.06 \; m_{\rm Pl}^2$.)
This result is quite remarkable since there exist longstanding speculations in the literature about
a possible de Sitter core of realistic black holes \cite{frol}. 
Those speculations were based upon purely phenomenological considerations and no derivation from
first principles has been given so far.\footnote{However, two-dimensional dilaton 
gravity has been shown \cite{strominger} to contain nonsingular quantum black holes asymptotic
to de Sitter space.} Instead, if the renormalization group improved metric  is 
reliable also at very short distances, the de Sitter core and in particular the 
regularity of the metric at $r=0$ is an automatic consequence. The validity of the improved
$f(r)$ for $r\rightarrow 0$ will be discussed in detail in section VIII.

The de Sitter metric (\ref{4.24}) has a ``cosmological'' horizon at 
$r_{\rm dS}=\sqrt{3/\Lambda_{\rm eff}}$. This value coincides precisely with the approximate
$r_{-}$ of eq.(\ref{4.22}).
The asymptotic de Sitter form (\ref{4.24})  is obtained only if $\gamma>0$. For
$\gamma =0$ the expansion starts with a term linear in $r$:
\be\label{4.26}
f(r)=1-{2 M\over \tilde{\omega}}\; r+O(r^2)\hspace*{2cm} (\gamma=0)
\ee
This spacetime is {\it not} regular at $r=0$, there remains a curvature singularity 
at the origin. We shall come back to this point in section VIII.
\subsection{The special case $\gamma=0$}
While close to $r=0$ (where the use of our improved $f(r)$ is anyhow questionable) the
physics implied by the quantum corrected metric strongly depends on the parameter $\gamma$,
the essential features of the spacetime related to larger distances are fairly insensitive
to the value of $\gamma$. In particular, it is easy to see that the general pattern of horizons
(two, one, or no horizon, their $M$-dependence, etc.) is qualitatively the same for all values of
$\gamma$. Even $\gamma=0$ gives the same general picture as the preferred value $\gamma=9/2$.
Therefore some of the calculations in the following sections will be performed for $\gamma=0$
which simplifies the algebra and leads to much more transparent results. For $\gamma=0$
we have, for instance,
\ba\label{4.27}
&&x_{\pm}=r_{\pm}/G_0 M=1\pm\sqrt{1-\Omega}\\[2mm]
&&\Omega_{\rm cr}=1, \;\;\;\; x_{\rm cr} = 1\\[2mm]\label{4.29}
&&M_{\rm cr}=\sqrt{\tilde{\omega}/G_0}\\[2mm]
&&r_{\rm cr}=\sqrt{\tilde{\omega} G_0}
\ea
It is amusing to see that the explicit formula for the location of the horizons, 
eq.(\ref{4.27}), coincides exactly with the corresponding expression for the 
\RN  black hole, eq.(\ref{4.17}). Note also that because of 
(\ref{4.29}) the parameter $\Omega$ can be interpreted as the ratio
\be\label{4.31}
\Omega = {M_{\rm cr}^2\over M^2}\hspace*{2cm} (\gamma=0)
\ee
\subsection{Geodesics and causal structure}
The global structure of our black hole spacetime is quite similar to  
the one of the \RN charged black hole. In particular,
the $r=0$ hypersurface is timelike now. 
The Penrose diagram of the spacetime is shown in Fig.(\ref{fig4})
for $M>M_{\rm cr}$. It is clear from the location of the horizons that we
can distinguish the following main regions:

\noindent
\hspace*{4cm} I and V : ~~~$r_{+}<r<\infty$\\
\hspace*{4cm} II and IV:  ~~~$r_{-}<r<r_{+}$\\
\hspace*{4cm} III and III':~~~$0<r<r_{-}$ 

The features of the motion in such a spacetime are
particularly evident if we consider a test particle 
which moves radially on a timelike geodesic. 
The equations of motion are given by 
\ba\label{5.10}
&&{dr\over d\tau}=\pm \left ({\cal E}^2-f(r)\right )^{1/2}\\[2mm]
&&{dv\over d\tau}=f(r)^{-1}\left ({\cal E}\pm({\cal E}^2-f(r))^{1/2}\right)\label{5.11}
\ea
where we have used Eddington-Finkelstein coordinates with $v$ being the advanced 
time coordinate.
Furthermore ${\cal E}$ denotes the constant of motion associated with the 
timelike Killing vector field $\xi^\mu = \delta^\mu_v$, 
\be\label{5.12}
{\cal E}=-\xi_\mu u^\mu
\ee
where $u^\mu$ is the four-velocity of a static observer. The choice of the sign in eq.(\ref{5.10})
and (\ref{5.11}) depends upon whether the test particle is travelling on a path 
of decreasing ($-$)  or increasing ($+$) radius $r$. 
From eq.(\ref{5.10}) we deduce that the proper acceleration is 
\be\label{5.13}
{d^2 r\over d\tau ^2}=-{1\over 2}{\partial f(r)\over \partial r}=
-{M G_0 r \; (r^3-\tilde{\omega}G_0 r-2\tilde{\omega}G_0^2
\gamma M)\over (r^3+\tilde{\omega} G_0\;[r+\gamma G_0 M])^2}
\ee
where from one sees that the radial motion is ruled by a Newton-type equation of motion with 
respect to the proper time $\tau$. It contains  the potential function 
$\Phi(r)={1\over 2}f(r)$ with the properties
$\Phi(0)=\Phi(\infty)={1\over 2}\equiv\Phi_{\rm max}$ and
$\Phi_{\rm min}<0$.
If we identify the ``energy'' of the motion with 
$\bar{{\cal E}}= {\cal E}^2/2$ we have from (\ref{5.10}) 
\be\label{5.13bis}
{1\over 2}\; \dot{r}^2+\Phi(r) = \bar{{\cal E}}
\ee
\begin{figure}
\hbox to\hsize{\hss\epsfxsize=4.5cm\epsfbox{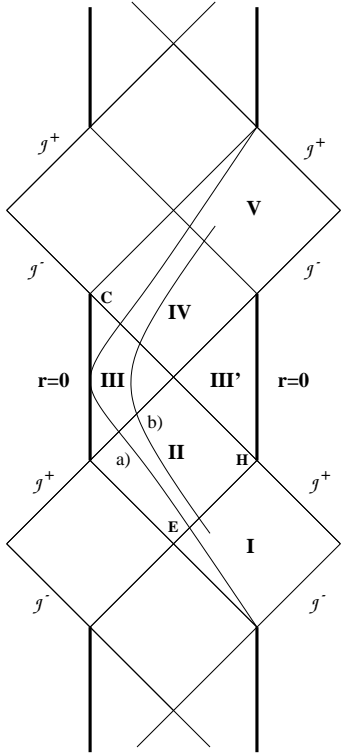}\hss}
\caption{Penrose conformal diagram of the quantum black hole
spacetime
\label{fig4}}
\end{figure}

It is thus possible to discuss the radial motion by the help of simple mechanical arguments
referring to Fig.(\ref{fig2}). In particular we use (\ref{5.13bis}) in order to 
determine the inflection points, {\it i.e.} zero-velocity configurations where
$\Phi(r)=\bar{{\cal E}}$, and where the sign in (\ref{5.10}) and (\ref{5.11}) has to be
changed.

There are basically three types of motions depending on the value of $\bar{{\cal E}}$:
\begin{description}
\item{({\it i})}  ~~$\bar{{\cal E}}> \Phi_{\rm max}$. The motion is unbounded. 
A free falling test particle that starts its motion
in region I, crosses the event horizon (EH in Fig.(\ref{fig4})) and 
eventually reaches $r=0$ in region III in a finite amount of proper time, with non-zero velocity 
and finite proper acceleration. It would thus cross the inner horizon (CH in Fig.(\ref{fig4}))
and continue its journey in regions IV and V. 
This is for instance the path $a)$ in Fig.(\ref{fig4}). 
It should be noted that this behavior is unlike the  
\RN one. In that case $\Phi(0)=\infty$ and there is always an inflection point in
region III. The particle is thus bounced away from the \RN central 
singularity at some non-zero value of the radius, before continuing its motion in region IV. 
\item{({\it ii})} ~$\bar{{\cal E}}\in [0,\Phi_{\rm max}]$. The motion is bounded. Starting in
region I it evolves into regions II and III and it eventually continues in region IV and V.
Let us first consider the case $\bar{{\cal E}}< \Phi_{\rm max}$. Then there is an inflection
point in region III and $r=0$ is avoided. A further inflection point is present in region V
where the trajectory reaches the same initial conditions as in region I.
The situation is shown in Fig.(\ref{fig4}) with the path b). 
If $\bar{{\cal E}}= \Phi_{\rm max}$ the inflection point is at $r=0$. If $\gamma\not =0$ this
is also an equilibrium point since, as it follows from (\ref{5.13bis}), the proper acceleration
is zero, $\Phi'(0)=0$. The particle reaches the center in an infinite amount of proper time. 
If $\gamma = 0$ the proper acceleration at $r=0$ is not zero, 
$\Phi'(0)=-\tilde{\omega} /M$, and $r=0$ is not an equilibrium configuration. Close to the origin 
the particle feels a repulsive force of strength $\tilde{\omega}/M$.
\item{({\it iii})} ~$\bar{{\cal E}} \in [\Phi_{\rm min},0]$.
The motion is bounded. It starts in region II where it has two inflection points
and it continues indefinitely in this region. 
\end{description}
It would be possible to study along similar lines spacelike and null geodesics as well as
the $M\leq M_{\rm cr}$ cases, but we shall not present this
analysis here. 
\renewcommand{\theequation}{5.\arabic{equation}}
\setcounter{equation}{0}
\section{Effective matter interpretation and Energy conditions }
Let us suppose that our quantum black hole has been generated 
by an ``effective'' matter fluid that simulates the effect of the quantum fluctuations of the metric.
We assume that this coupled gravity-matter system satisfies the conventional Einstein equations
$G_{\mu\nu} = 8\pi G_0 T_{\mu\nu}$. The stress-energy tensor of a (not necessarily classical) 
perfect fluid with the symmetries of our spacetime reads
\be\label{5.1}
{T^{\alpha}}_{\beta}={\rm diag}\;(-\rho , P_r, P_{\perp}, P_{\perp})
\ee
It is thus possible to use the Einstein equations in order to {\it define} the components
of ${T^{\alpha}}_{\beta}$
in the following way
\ba\label{5.2}
&&-8\pi G_0 \;\rho = G_t^t = G_r^r\\[2mm]
&&8\pi G_0 \;P_\perp = G_\theta^\theta \label{5.2a}
\ea
Here $G_{\mu\nu}$ is the Einstein tensor of the metric (\ref{4.1}) with
(\ref{4.4}). A straightforward calculation shows that 
\ba\label{5.3}
&&\rho=-P_r= {1\over 4\pi}\;
{M G_0\tilde{\omega}\;(2r+3\gamma G_0 M)\over 
(r^3+\tilde{\omega}G_0\;[r+\gamma G_0 M])^2}\\[2mm]
&&P_{\perp}={1\over 4\pi}\;
{M G_0\tilde{\omega}\; (3r^4+6r^3G_0\gamma M-3\tilde{\omega}G_0^3\gamma^2 M^2
-r^2\tilde{\omega}G_0-3\tilde{\omega}G^2r\gamma M)
\over 
(r^3+\tilde{\omega}G_0\;[r+\gamma G_0 M])^3}
\ea
The total energy outside a given radius $r$,
\be\label{5.4}
E(r) = 4\pi \int_r^\infty \rho(r') \; r'^2\;dr'
\ee
is given by 
\be\label{5.4bis}
E(r) = {\tilde{\omega} G_0 M \;
(r+\gamma G_0 M) \over r^3+\tilde{\omega}G_0\;[r+\gamma G_0 M]}
\ee
This quantity is finite and positive definite for any value of $\gamma$. 
In particular we find the surprisingly simple result  
\be\label{5.6}
E_{\rm tot}\equiv E(0) = M
\ee
which identifies the total energy of the fluid with the mass of the black hole.

It is not difficult to realize that this ``magic'' equality is a consequence of 
the boundary conditions set at 
spatial infinity, $f=1-2G_0M/r +O(1/r^2)$, and of the behavior of $f$ at $r=0$. For 
{\it every} metric of the form (\ref{4.1}), with an arbitrary function $f(r)$, 
the definitions (\ref{5.2}) and (\ref{5.4}) lead to the expression
\be\label{5.7}
E(r) = -{1\over G_0} \;\lim_{\hat{r}\rightarrow \infty}
\int^{\hat{r}}_r ds [(sf(s))'-1]=M+{r\over 2G_0}(f(r)-1)
\ee
Obviously $E(0)$ equals $M$ if $rf(r)\rightarrow 0$ when $r \rightarrow 0$, and this is always
satisfied in our model, for any value of $\gamma$. 
Note that in ordinary Schwarzschild spacetimes with ADM-mass $M$, since $r-rf(r)=2G_0M$,
it follows that $E_{\rm tot}=0$. In our picture the quantum effects can be interpreted 
as a non-zero $\rho$ and $P_\perp$. It is thus remarkable that their global contribution
is exactly equal to the total mass of the spacetime. 

For $M<M_{\rm cr}$ we have seen from the discussion of section III
that no horizon is present and that, contrary to what happens in classical
\RN  spacetimes for $\Omega_{\rm RN}>1$, no naked singularity occurs
(for $\gamma>0$).  The spacetime, in this case, resembles a soliton-like
particle with a  planckian rest mass given by (\ref{5.6}).
The energy of the fluid is then localized in a cloud around $r=0$ 
with $\partial_r E(r)<0$ always, and $E(\infty)=0$. 

It is possible to show that our ``effective'' fluid does not meet all the 
requirements in order to be considered a classical fluid. In fact, it violates the 
dominant energy condition in some regions, depending on the values of $\gamma$. 
In particular, the condition $P_{\perp}-\rho\leq 0$ is not always satisfied 
since  it amounts to 
\be\label{5.7a}
r^4+3r^3\gamma G_0 M-3r^2\tilde{\omega}G_0 -8r\tilde{\omega}\gamma G_0^2M \leq 0
\ee
which does not hold for some interval of values of the radial coordinate. 
For instance, for $\gamma=0$ the left hand side of 
(\ref{5.7a}) is positive when $r\leq \sqrt{3\tilde{\omega} G_0}$.

In the improved black hole spacetimes there are also  ``zero gravity'' hypersurfaces, where  
the Weyl and Ricci curvature are zero,  in analogy to what has been found  
in \cite{dimnikova} in a phenomenological model with a de Sitter core.
In fact the ``Coulombian'' component of the Weyl tensor can be shown to be
\be\label{5.8}
\Psi_2 = -{1\over 3} {M G_0 r\; (3r^5 -r^3 \tilde{\omega} G_0
-6r^2\;\tilde{\omega}\gamma G_0^2 M 
- \tilde{\omega}^2\gamma G_0^3 M )\over
(r^3+\tilde{\omega}\;G_0 [r+\gamma G_0 M])^3}
\ee 
It should be noted that the Weyl curvature is regular at $r=0$
where it is always zero. Similarly, the Ricci scalar reads
\be\label{5.9}
R=-{4\tilde{\omega}\;G_0M\;(r^4+3r^3\gamma G_0 M-3r^2\tilde{\omega} G_0
-8 r\tilde{\omega}\gamma G_0^2 M-6\tilde{\omega}\gamma^2 G_0^3 M^3)
\over (r^3+\tilde{\omega}\;G_0 [r+\gamma G_0 M])^3}
\ee
In general the location of zero-curvature hypersurfaces depends on the value of $\gamma$
and of the black hole mass. In the case of $\gamma=0$ one sees from the above expressions 
that at the radii $r=\sqrt{\tilde{\omega}G_0/3}$ and 
$r=\sqrt{3\tilde{\omega}G_0}$ one or the other of the two main scalar curvature invariants of our spacetime
is zero. In particular, for $M<M_{\rm cr}$ and $\gamma =0$ 
the radius 
\be\label{5.9bis}
r=\sqrt{3\tilde{\omega}\;G_0}
\ee
can be thought of as the characteristic length of the particle-like soliton 
structure arising from the renormalization group improvement of the spacetime of
a nearly planckian black hole. 
\renewcommand{\theequation}{6.\arabic{equation}}
\setcounter{equation}{0}
\section{hawking temperature and black hole evaporation}
\subsection{The Euclidean manifold}
Let us consider Lorentzian black hole metrics of the type (\ref{4.1}) with an essentially 
arbitrary function $f(r)$. For the time being we only assume that $f$ has a simple 
zero at some $r_{+}$, $(f(r_{+})=0, f'(r_{+})\not= 0)$ and that it increases monotonically 
from zero to $f(\infty)=1$ for $r>r_{+}$. The behavior of $f(r)$ for $r<r_{+}$ will not matter 
in the following. There exists a standard method for associating a Euclidean black hole 
spacetime to metrics of this type \cite{euclidean}. The first step is to perform a ``Wick rotation''
by setting $t=-i\tau$ and taking $\tau$ real:
\be\label{6.1}
ds^2_{E}=f(r)\; d\tau^2+f(r)^{-1}\; dr^2+r^2d\Omega^2
\ee
This line element defines a Euclidean metric on the manifold coordinatized by 
$(\tau,r,\theta,\phi)$ with $r>r_{+}$ where $ds^2_{E}$ is positive definite. In order to 
investigate the properties of this manifold we trade $r$ for a new coordinate $\rho$ defined by
\be\label{6.2}
\rho = {1\over 2\pi}\;\beta_{\rm BH}\sqrt{f(r)}
\ee
with the constant
\be\label{6.3}
\beta_{\rm BH}\equiv {4\pi\over f'(r_{+})}
\ee
The new coordinate ranges from $\rho=0$ to $\rho = \beta_{\rm BH}/2\pi$ corresponding 
to $r=r_{+}$ and $r\rightarrow\infty$, respectively. Thus the line element becomes
\be\label{6.4}
ds^2_{E} = \rho^2\; \left ({2\pi\over \beta_{\rm BH}}\right )^{2}\; d\tau^2
+\left [ {f'(r_{+})\over f'(r(\rho))}\right ]^2\; d\rho^2 +r(\rho)^2 \; d\Omega^2
\ee
where $r$ is a function of $\rho$ now. Close to the horizon $(r=r_{+}$ or $\rho=0)$ 
this metric simplifies considerably:
\be\label{6.5}
ds^2_{E}\approx \rho^2 \; d\hat{\tau}^2+d\rho^2+r_{+}^2 \; d\Omega^2
\ee
In writing down (\ref{6.5}) we introduced the rescaled Euclidean time
\be\label{6.6}
\hat{\tau}\equiv {2\pi\over \beta_{\rm BH}}\; \tau
\ee
Leaving aside the $r_{+}^2 d\Omega^2$-term for a moment, we see that (\ref{6.5}) looks like
the metric of a 2-dimensional Euclidean plane written in polar coordinates 
$\hat{\tau}$ and $\rho$. For this to be actually true, $\hat{\tau}$ must be considered an angular 
variable with period $2\pi$. If $\hat\tau$ is periodic with a period different from $2\pi$,
the space has a conical singularity at $\rho=0$. In order to avoid this singularity we require
the unrescaled time $\tau$ to be an angle-like variable with period $\beta_{\rm BH}$,
$\tau\in [0,\beta_{\rm BH}]$. With this periodic identification, eq.(\ref{6.4}) defines 
a Euclidean metric on what is referred to as the Euclidean black hole manifold. It has the 
topology of $R^2\times S^2$. 

If we put quantized matter fields on the Euclidean black hole spacetime their Green's 
functions inherit the periodicity in the time direction. Thus they appear to be thermal Green's 
functions with the temperature given by
\be\label{6.7}
T_{\rm BH}=\beta_{\rm BH}^{-1}={1\over 4\pi} f'(r_{+})
\ee
This is the Bekenstein-Hawking temperature for the general class of black holes with metrics
of the type (\ref{4.1}). 
\subsection{Temperature and specific heat}
Eq.(\ref{6.7}) is an essentially ``kinematic'' statement and 
its derivation does not assume any specific form of the field equations for the metric. 
Hence we may apply it to the renormalization group improved Schwarzschild metric and investigate
how the quantum gravity  effects modify the Hawking temperature. By differentiating 
eq.(\ref{4.4}) we find 
\be\label{6.8}
T_{\rm BH}(M)={1\over 8\pi G_0 M}\; \left [1-{\Omega\over x_{+}^2}-
{2\; \gamma\; \Omega\over x_{+}^3}\right ]
\ee
where $x_{+}$ and $\Omega$ are considered functions of $M$. When we switch off quantum gravity 
$(\tilde{\omega}=0)$ or look at very heavy black holes $(M\rightarrow \infty)$
we have $\Omega = 0$ and recover the classical result 
\be\label{6.9}
T_{\rm BH}^{\rm class}(M)={1\over 8\pi G_0 M}
\ee
(In the present context, the term ``classical'' refers to the usual ``semiclassical''
treatment with quantized matter fields on a classical geometry.) Obviously the quantum 
corrected Hawking temperature is always smaller than the classical one: 
$T_{\rm BH}<T_{\rm BH}^{\rm class}$. 

In order to be more explicit we continue our investigation for the special value $\gamma=0$.
Using (\ref{4.27}) and (\ref{4.29}) we obtain 
\ba
T_{\rm BH}(M) && = {1\over 4\pi G_0 M}\;{\sqrt{1-\Omega}\over 1
+\sqrt{1-\Omega}}\\[2mm]\label{6.10}
&& ={1\over 4\pi G_0 M_{\rm cr}}\; {\sqrt{\Omega(1-\Omega)}\over 1+\sqrt{1-\Omega}}\label{6.11}
\ea
with $\Omega= M^2_{\rm cr}/M^2$. Eq.(\ref{6.10}) is quite similar, though not identical to 
the corresponding expression for the temperature of the \RN black hole:
\be\label{6.12}
T_{\rm BH}^{\rm RN}={1\over 2\pi G_0 M}\; {\sqrt{1-\Omega_{\rm RN}}\over 
(1+\sqrt{1-\Omega_{\rm RN}})^2}
\ee

\begin{figure}[ht]
\hbox to\hsize{\hss\epsfxsize=12cm\epsfbox{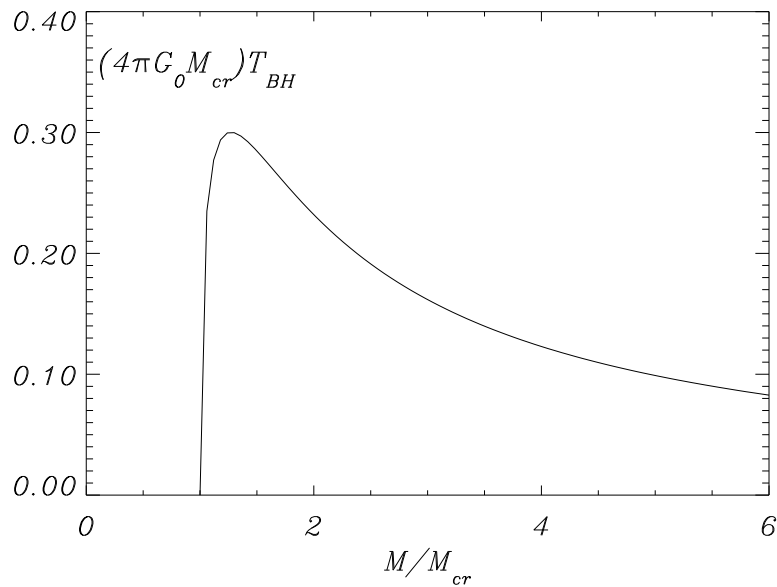}\hss}
\caption{
The Hawking temperature of the quantum black hole (multiplied by 
$4\pi G_0 M_{\rm cr}$) as a function of $M/M_{\rm cr}$. The maximum 
temperature is reached for $\widetilde{M}_{\rm cr}\approx 1.27 M_{\rm cr}$.
\label{fig5}}
\end{figure}

The large mass expansion of $T_{\rm BH}$ reads
\be\label{6.13}
T_{\rm BH}(M)={1\over 8\pi G_0 M}
\left [1-{1\over 4}\left ({M_{\rm cr}\over M}\right )^2-{1\over 8}
\left ( {M_{\rm cr}\over M} \right ) ^4 +O(M^{-6})\right ]
\ee
with $M^2_{\rm cr}=\tilde{\omega}/G_0$. Probably the first few terms of this series 
are a rather precise prediction of our method because they correspond to a spacetime which is only 
very weakly distorted by quantum effects.

Let us look at what happens when $M$ approaches $M_{\rm cr}$ from above. We set
\be\label{6.14}
\Omega=\Omega_{\rm cr}-\epsilon= 1-\epsilon
\ee
and study the limit $\epsilon\rightarrow 0^{+}$. Note that because  
$\Omega=M_{\rm cr}^2/ M^2$ for $\gamma=0$,
\be\label{6.15}
\epsilon= 1-{M^2_{\rm cr}\over M}
\ee
Expanding (\ref{6.11}) yields
\be\label{6.16}
T_{\rm BH}(M)=
{\sqrt{\epsilon}\over 4\pi G_0 M_{\rm cr}}\;\left [ 1-\sqrt{\epsilon}+{1\over 2}\epsilon+
O(\epsilon^{3/2})\right ]
\ee
or, to lowest order,
\be\label{6.17}
T_{\rm BH}(M)=
{1\over 4\pi \tilde{\omega}}\; \sqrt{M^2-M_{\rm cr}}+O(M^2-M_{\rm cr})
\ee
We see that $T_{\rm BH}$ vanishes as $M$ approaches its critical value $M_{\rm cr}$.
This conclusion is true for any value of $\gamma$. In fact, as $M\searrow M_{\rm cr}$ 
the simple zero of $f$ at $r_{+}$ tends to become a double zero, 
{\it i.e.} $f'(r_{+})\rightarrow 0$, and therefore $T_{\rm BH}\rightarrow 0$ by 
eq.(\ref{6.7}). We emphasize, however, that the statement 
\be\label{6.18}
T_{\rm BH}(M_{\rm cr}) = 0
\ee
should always be understood in the sense of a limit $M\searrow M_{\rm cr}$
because strictly speaking the above derivation of the Hawking 
temperature does not apply for the critical (extremal) black hole with $M$ exactly equal to 
$M_{\rm cr}$. 

In Fig.(\ref{fig5}) the Hawking temperature is plotted for the full range of mass values. For large values 
of $M$ we recover the classical $1/M$-decay, and for $M>M_{\rm cr}$ the temperature increases with 
$M$. The Hawking temperature reaches its maximum for a certain mass $\widetilde{M}_{\rm cr}
>M_{\rm cr}$ which plays the role of another ``critical'' temperature (see below). 
By definition, 
\be\label{6.19}
{d T_{\rm BH}\over d M} (\widetilde{M}_{\rm cr})=0 
\ee
The $\Omega$-value related to $\widetilde{M}_{\rm cr}$ will be denoted 
$\widetilde{\Omega}_{\rm cr}$; for $\gamma=0$ it reads 
\be\label{6.20}
\widetilde{\Omega}_{\rm cr}=\left ({M_{\rm cr}\over \widetilde{M}_{\rm cr}}\right)^2
\ee
Differentiating (\ref{6.11}) leads to $\widetilde{\Omega}_{\rm cr}=(\sqrt{5}-1)/2\approx 0.62$
which yields $\widetilde{M}_{\rm cr}=M_{\rm cr}\widetilde{\Omega}_{\rm cr}^{-1/2}
\approx 1.27 M_{\rm cr}$. The maximum at $\widetilde{M}_{\rm cr}$ is surprisingly 
close to $M_{\rm cr}$ so that the drop from the peak value of $T_{\rm BH}$ down to zero is rather
steep.
\begin{figure}[ht]
\hbox to\hsize{\hss\epsfxsize=12cm\epsfbox{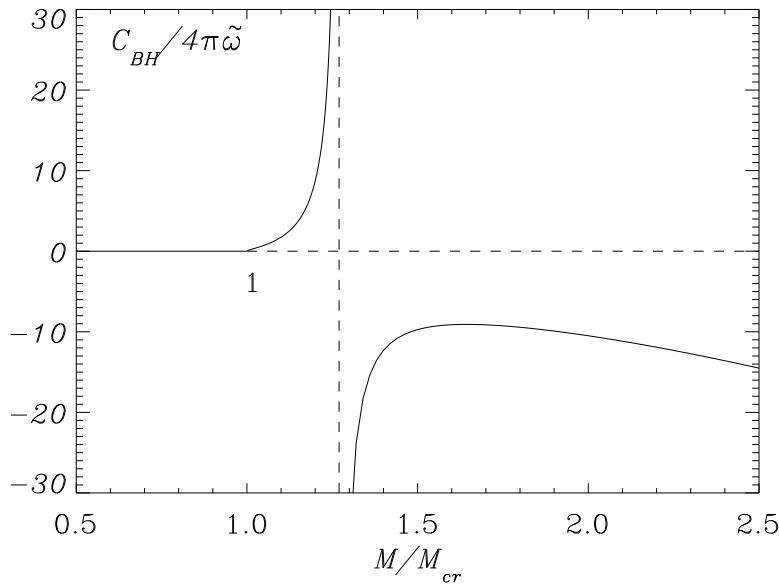}\hss}
\caption{
The specific heat $C_{\rm BH}$ in units of $4\pi \tilde{\omega}$ as a function of
$M/M_{\rm cr}$. The singularity occurs at $\widetilde{M}_{\rm cr}/M_{\rm cr}\approx 1.27$. 
\label{fig6}}
\end{figure}

Even though we do not have a full statistical mechanical formalism with a partition function
and a free energy functional at our disposal, eq.(\ref{5.6}) 
suggests to identify the internal energy $U$ of the black hole with  its total mass $M$. 
Then the standard thermodynamical relation $C_{V}=(\partial U/\partial T)_V$ amounts to the 
following definition for the specific heat capacity of the black hole:
\be\label{6.21}
C_{\rm BH}= {d M\over d T_{\rm BH}} = \left ({d T_{\rm BH} \over d M}\right )^{-1}
\ee
From eq.(\ref{6.11}) we obtain 
\be\label{6.22}
C_{\rm BH}=
-4\pi\tilde{\omega}\; {(1-\Omega)\; [1+\sqrt{1-\Omega}]^2\over\Omega[\Omega^2
+(1-2\Omega)(1+\sqrt{1-\Omega})]}
\ee
The specific heat $C_{\rm BH}$ is negative for $M>\widetilde{M}_{\rm cr}$
and becomes positive for $M_{\rm cr}<M<\widetilde{M}_{\rm cr}$. It has a singularity at
$M=\widetilde{M_{\rm cr}}$ which signals a kind of phase transition at this value of the mass. 
In Fig.(\ref{fig6}), $C_{\rm BH}$ is shown as a function of $M$.  For very heavy black holes one has
\be\label{6.23}
C_{\rm BH}=
-8\pi G_0 M^2 \left [1+{3\over 4}\left ({M_{\rm cr}\over M}\right )^2 +
{19\over 16}\left ({M_{\rm cr}\over M} \right )^4
+O(M^{-6})\right ]
\ee
In the limit $M\rightarrow \infty$ we recover the classical value 
$C_{\rm BH}^{\rm class}= -8\pi G_0 M^2$, and we observe that the leading quantum corrections
make the already negative specific heat even more negative\footnote{ This is the same 
tendency as in the Weyl-gravity model of ref.\cite{hasslacher}, for instance.}. 
In the limit $M\searrow M_{\rm cr}$ the specific heat vanishes according to 
\be\label{6.24}
C_{\rm BH}= 
4\pi\tilde{\omega} \; \sqrt{\epsilon}\; \Big [1+2\sqrt{\epsilon}+4\epsilon
+O(\epsilon^{3/2})\Big ]=
4\pi\tilde{\omega}\; \sqrt{1-{M_{\rm cr}^2\over M^2}}\; +\cdots
\ee
\subsection{Stopping the evaporation process}
From our result for the mass dependence of the Bekenstein-Hawking temperature the following 
scenario for the black hole evaporation with the leading quantum correction included emerges.
As long as the black hole is very heavy the classical relation 
$T_{\rm BH}\propto 1/M$ is approximately valid. The black hole radiates off energy,
thereby lowers its mass and increases its temperature. This tendency is counteracted by the quantum 
effects. The actual temperature stays always below $T^{\rm class}_{\rm BH}$. Once the mass is as 
small as $\widetilde{M}_{\rm cr}$, the temperature reaches its maximum value 
$T_{\rm BH}(\widetilde{M}_{\rm cr})$. For even smaller masses it drops very rapidly and it 
vanishes once $M$ has reached its critical mass $M_{\rm cr}$, which is of the order of 
$m_{\rm Pl}$.

In the classical picture based upon $T_{\rm BH}^{\rm class}\propto 1/M$ the
black hole becomes continuously hotter during the evaporation process. In the above scenario, on 
the other hand, its temperature  never exceeds $T_{\rm BH}(\widetilde{M}_{\rm cr})$, and the
evaporation process comes to a complete halt when the mass has reached $M_{\rm cr}$.
This suggests that the critical (or extremal) black hole with $M=M_{\rm cr}$ could be the final state
of the evaporation of a Schwarzschild black hole. If stable, the critical black hole 
would indeed constitute a Planck-size remnant of burnt-out macroscopic black holes.
It is ``cold'' in the sense that $\lim_{M\searrow M_{\rm cr}} T_{\rm BH}(M)=0$, so that 
it is stable at least against the classical Hawking radiation mechanism as we know it.

It is interesting to see how long it takes a black hole with the initial mass $M_{\rm i}$  to reduce its
mass to some final value $M_{\rm f}$ via Hawking radiation. Stefan's law provides us with 
a rough estimate of the radiation power.  The mass-loss per unit proper time of an infinitely 
far away, static observer is approximately given by
\be\label{6.25}
-{d M\over d t} = \sigma\; {\cal A}(M)\; T_{\rm BH}(M)^4
\ee
Here $\sigma$ is a constant and ${\cal A}=4\pi r_{+}^2$ is the area of the outer horizon:
\be\label{6.26}
{\cal A}(M)=8\pi G_0^2 M^2\;[1-{1\over 2}\Omega +\sqrt{1-\Omega}]
\ee
In the classical case the above differential equation becomes $-d M/dt \propto M^{-2}$.
It is easily integrated with the result that only a {\it finite} amount of time 
$t(M_{\rm i}\rightarrow 0)\propto M_{\rm i}^3$ is needed in order to completely radiate away the initial mass.
The problems such as the information paradox mentioned in the introduction are particularly severe
because the catastrophic end point of the evolution ($T_{\rm BH}\rightarrow \infty$) 
is reached within a finite time. 

Looking at the quantum black hole now, we assume that the initial mass $M_{\rm i}$ is already close to
$M_{\rm cr}$ so that we may use the approximation (\ref{6.17}) on the RHS of eq.(\ref{6.25}):
\be\label{6.27}
-{d M\over d t}={\sigma G_0 \over (4\pi\tilde{\omega})^3} \; (M^2-M_{\rm cr}^2)+\cdots
\ee
Obviously the radiation power decreases quickly as $M\searrow M_{\rm cr}$. Integrating 
(\ref{6.27}) yields for the time to go from $M_{\rm i}$ to $M_{\rm f}$:
\be\label{6.28}
t(M_{\rm i}\rightarrow M_{\rm f})=
16\pi^3\tilde{\omega}^2\sigma^{-1}\left ({1\over M_{\rm f}-M_{\rm cr}}
-{1\over M_{\rm i}-M_{\rm cr}}\right )
\ee
We see that this time diverges for $M_{\rm f}=M_{\rm cr}$, {\it i.e.} it takes an {\it infinitely long}
time to reduce the mass from any given $M_{\rm i}$ down to the critical mass. Clearly the reason is 
that, because of the $T^4$-behavior, the radiation power becomes very small when we approach the 
``cold'' critical black hole. In a certain sense, this result is a reflection of the third law of 
black hole thermodynamics which states that it is impossible to achieve an exactly vanishing surface 
gravity, {\it i.e. } $T=0$, by any physical process. 

The back reaction of the Hawking radiation on the metric is neglected in the above arguments. We
believe that most probably (contrary to the case of the classical Hawking black hole)
its inclusion would not lead to qualitative changes of the picture. The reason is that $dM/dt$ is
very small in both the early {\it and the late} stage of the evaporation process, 
and that in between
its value is bounded above.
\renewcommand{\theequation}{7.\arabic{equation}}
\setcounter{equation}{0}
\section{entropy of the quantum black hole}
One of the most intriguing aspects of black hole thermodynamics is the entropy associated with the 
horizon of a black hole.  It is one of the central but as to yet unresolved questions 
if and how this entropy can be interpreted within a ``microscopic'' statistical mechanics by 
counting the number of micro states which are unaccessible to our observation
\cite{wald}. Another important question is how the classical relation between the entropy and the 
surface area of the horizon
\be\label{7.1}
S_{\rm class}={{\cal A}_{\rm class}\over 4 G_0}
\ee
changes if quantum (gravity) effects are taken into account. 
\begin{figure}
\hbox to\hsize{\hss\epsfxsize=13cm\epsfbox{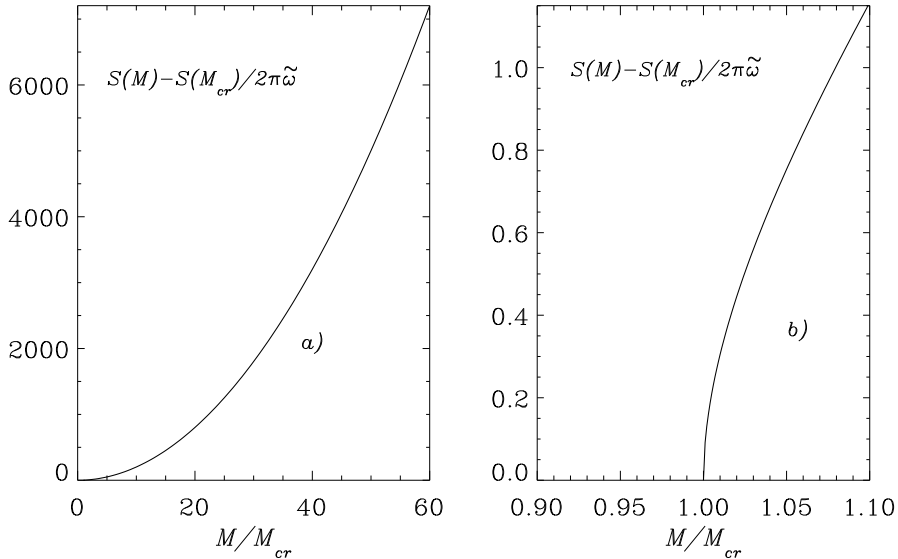}\hss}
\caption{
a) The entropy $S(M)-S(M_{\rm cr})$ in units of $2\pi\tilde{\omega}$
as a function of $M/M_{\rm cr}$.
b) The same function  for $M$ near $M_{\rm cr}$.
\label{fig7}}
\end{figure}

Our approach of renormalization group improving the Schwarzschild spacetime makes a definite
prediction for the quantum correction of the entropy.
The key ingredient is the function $T_{\rm BH}=T_{\rm BH}(M)$ which we obtained in section VI.
From general thermodynamics we know that the entropy $S=S(U,V,\cdots)$ 
satisfies $(\partial S/\partial U)_V=1/T$. In the present context we identify the energy $U$
with the mass $M$, and since the volume dependence plays no role $S=S(M)$ satisfies
$dS / dM = 1/T_{\rm BH}(M)$. Upon integration we have
\be\label{7.2}
S(M)- S(M_{\rm cr}) = \int _{M_{\rm cr}}^M {d M'\over T_{\rm BH}(M')}
\ee
where the reference point was chosen to be the critical mass. For simplicity we continue the
analysis for $\gamma = 0$; inserting the corresponding Hawking temperature (\ref{6.11}) into
(\ref{7.2}) we obtain
\be\label{7.3}
S(M)-S(M_{\rm cr}) = 2\pi \tilde{\omega}\int^1_{M_{\rm cr}^2/ M^2} \;
{d\Omega\over \Omega^2}\; \left [1+{1\over \sqrt{1-\Omega}}\right ] 
\ee
The integral yields for $M\geq M_{\rm cr}$ 
\be\label{7.4}
S(M)-S(M_{\rm cr}) = 2\pi\tilde{\omega}\;
\left [ \Omega^{-1}\sqrt{1-\Omega}\; (1+\sqrt{1-\Omega})
+ {\rm artanh}\sqrt{1-\Omega} \right ]
\ee
with $\Omega\equiv M_{\rm cr}^2 / M^2$ on the RHS of (\ref{7.4}). Eq.(\ref{7.4}) is our prediction
for the quantum corrected entropy of the black hole geometry. Its large-$M$ expansion reads
\be\label{7.5}
S(M)-S(M_{\rm cr})=
{{\cal A}_{\rm class}\over 4 G_0} + 2\pi\tilde{\omega}\;
\left [ \ln\left({2 M\over M_{\rm cr}}\right) - {3\over 2} -{3\over 8}
\left({M_{\rm cr}\over M}\right)^2 -{5\over 32}\left({M_{\rm cr}\over M}\right)^4+O(M^{-6})\right ]
\ee
with the classical area ${\cal A}_{\rm class}= 4\pi (2G_0 M)^2$. For very heavy black holes we 
recover the classical entropy $S_{\rm class}$ as the difference of $S(M)$ and the integration 
constant $S(M_{\rm cr})$ whose value remains undetermined here. The leading quantum correction
is proportional to $\ln (M)$. Remarkably, very similar $\ln(M)$-terms had been found with rather 
different methods \cite{frol2}. While some of the earlier results were plagued by the presence
of numerically undefined cutoffs, eq.(\ref{7.4}) is perfectly finite. When $M$ approaches 
$M_{\rm cr}$ from above, the entropy difference displays a square-root behavior:
\be\label{7.6}
S(M)-S(M_{\rm cr})=
4\pi\tilde{\omega}\;\sqrt{\epsilon}\;[1+{1\over 2}\sqrt{\epsilon}+O(\epsilon)]
\ee
In Fig.(\ref{fig7}) the entropy is shown as function of $M$.

The above calculation of $S(M)$ was within the framework of ``phenomenological''
thermodynamics. For an attempt at interpreting it within an underlying statistical mechanics
we refer to the Appendix.
\renewcommand{\theequation}{8.\arabic{equation}}
\setcounter{equation}{0}
\section{is there a curvature singularity at {\normalsize{ \lowercase{$\bf r=0$}}} ?}
We saw already that for $r\rightarrow 0$ the renormalization group improved black hole metric
approaches that of de Sitter space. The quantum black hole seems to have a ``de Sitter core'' of a 
similar type as the regular black holes which were introduced in 
ref.\cite{dimnikova} on a phenomenological basis.
This means in particular that the quantum corrected spacetime
is completely regular, {\it i.e.} contrary to the ordinary Schwarzschild black hole it is free
from any curvature singularity. However, because the classical and the quantum geometries are very
different for $r\rightarrow 0$ and the quantum effects play a dominant role there, 
it seems problematic to describe
a possibly regular core as an ``improvement'' of the singular Schwarzschild spacetime. 
Therefore some comments concerning the applicability of our approximation 
at very small distances are appropriate.

The regularity of the improved metric comes about because the $1/r$-behavior 
of $f_{\rm class}=1-2G_0 M/r$ is tamed by a very fast vanishing of the Newton constant at small 
distances. Close to the core of the black hole we are in the regime where the running of 
$G(k)$ is governed by the UV-fixed point, $G(k)\approx 1/\omega k^2$, so that the position-dependent
Newton constant is approximately given by
\be\label{8.1}
G(r) \approx \tilde{\omega}^{-1}\; d(r)^2
\ee
It is important to keep in mind that the distance function $d(r)$ depends on the classical
metric which we are going to improve. In section III we started from the 
Schwarzschild background and found that for all sensible curves
${\cal C}$, 
\be\label{8.2}
d_{\rm Sch}(r)\propto r^{3/2}
\ee
so that 
\be\label{8.3}
G_{\rm Sch}(r)\propto r^3
\ee
Taking (\ref{8.3}) literally means that the improved $f=1-2 G(r)M/r$ is of the de Sitter form
$1-({\rm const})\; r^2$ for $r\rightarrow 0$. 

However, if the actual quantum geometry really was de Sitter, there is no point in evaluating
$d(r)$ for the Schwarzschild background. In fact, if we calculate $d(r)$ for the de Sitter metric
the asymptotic behavior is different:
\be\label{8.4}
d_{\rm dS}(r) \approx r
\ee
Incidentally, this is precisely the $d$-function which obtains by setting $\gamma=0$ in 
eq.(\ref{3.13}). Eq.(\ref{8.4}) entails that the Newton constant vanishes more slowly than in 
(\ref{8.3}):
\be\label{8.5}
G_{\rm dS}(r)\propto r^2
\ee
Inserting (\ref{8.5}) into $f_{\rm class}$ we obtain a lapse function which approaches $f=1$
only {\it linearly},
\be\label{8.6}
f(r)=1-c\;r+O(r^2)
\ee
(Here $c$ is a constant.) 

The metric with an $f$-function of the general form 
\be\label{8.6a}
f(r) = 1-c \;r^\nu
\ee
where $c$ and $\nu$ are constant has the exact curvature invariants
\ba
R&&\;=\;c(\nu+1)(\nu+2)\;r^{\nu-2}\label{8.7a}\\[2mm]
R_{\mu\nu\rho\sigma}R^{\mu\nu\rho\sigma}&&\;=
\; c^2\; (\nu^4-2\nu^3+5\nu^2+4)\; r^{2\nu-4}\label{8.7b}\\[2mm]
C_{\mu\nu\rho\sigma}C^{\mu\nu\rho\sigma}&&\;= \;{c^2\over 144}\; (\nu-1)^2\; (\nu-2)^2\; 
\; r^{2\nu -4}\label{8.7c}
\ea
This means that the ``$G_{\rm dS}$-improved'' black hole of (\ref{8.6}) has a curvature
singularity at its center:
\ba
R&&\;=\;{6 c\over r}+\cdots\label{8.8a}\\[2mm]
R_{\mu\nu\rho\sigma}R^{\mu\nu\rho\sigma}&&\;=\;8 \;{ c^2\over r^2}+\cdots\label{8.8b}
\ea
Eq.(\ref{8.7c}) shows that the square of the Weyl tensor is regular for this metric.
Even if, contrary to the ``$G_{\rm Sch}$-improved'' spacetime, the  ``$G_{\rm dS}$-improved''
geometry is singular at the origin, it is much less singular than it was classically.
For the Schwarzschild metric one has
\be\label{8.9}
\Big ( R_{\mu\nu\rho\sigma}R^{\mu\nu\rho\sigma}\Big)_{\rm Sch}=
48{G_0^2M^2\over r^6}
\ee
with an additional factor of $1/r^4$ compared to (\ref{8.8b}). 

Within the present framework, 
we have no criterion for deciding whether the improvement $G_0\rightarrow G(r)$ should be done with 
$d_{\rm Sch}$, $d_{\rm dS}$, or the $d$-function of some unknown metric interpolating between 
Schwarzschild and de Sitter. This is a principal limitation of our approach. It appears plausible that
$f(r)\approx 1-c r^\nu$ for $r\rightarrow 0$ with the exponent $\nu$ somewhere in between the values
resulting from $d_{\rm Sch}$-improvement $(\nu=2)$ and $d_{\rm dS}$-improvement $(\nu=1)$. Except for 
$\nu=2$, the quantum black hole would have a curvature singularity at its center then. 
A reliable calculation of the exponent $\nu$ seems to be extremely difficult, though. 
Nevertheless it is probably a safe prediction that the central singularity is much weaker than  
its classical counterpart. The reason is that we found quantum gravity to be asymptotically free and that near the 
UV-fixed point $G(k)\propto 1/k^2$. In one way or another, this $k$ dependence must translate into
a ``switching off'' of the gravitational interaction at small distances.

The improvement with $d_{\rm dS}$ is equivalent to setting $\gamma=0$ in the formulas of 
the previous sections. While the cases $\gamma=0$ and $\gamma>0$ are qualitatively different
for $r\rightarrow 0$, we saw already that the other features of the quantum black holes
(horizons, Hawking radiation, entropy, etc.) are essentially the same in both cases.
\renewcommand{\theequation}{9.\arabic{equation}}
\setcounter{equation}{0}
\section{summary and conclusions}
In this paper we used the method of the renormalization group improvement
in order to obtain a qualitative understanding of the quantum gravitational 
effects in spherically symmetric black hole spacetimes. 

As far as the structure of horizons is concerned,  the quantum effects are small for 
very heavy black holes ($M\gg m_{\rm Pl}$). They have an event horizon at a radius 
$r_{+}$ which is close to, but always smaller than the Schwarzschild radius $2G_0M$. 
Decreasing the mass of the black hole the event horizon shrinks. There is also an inner (Cauchy)
horizon whose radius $r_{-}$ increases as $M$ decreases. For $M\rightarrow \infty$
it assumes its nonzero (if $\gamma\not = 0$) minimal value.  When $M$ equals the 
critical mass $M_{\rm cr}$ which is of the order of the Planck mass the two horizons
coincide. The near-horizon geometry of this critical black hole is that of $AdS_{2}
\times S^{2}$. For $M<M_{\rm cr}$ the spacetime has no horizon at all. 

While the exact fate of the singularity at $r=0$ cannot be decided  within our present approach,
we argued that either it is not present at all or it is at least much weaker than its classical counterpart.
In the first case the quantum spacetime has a smooth de Sitter core so that we are in accord with 
the cosmic censorship hypothesis even if $M<M_{\rm cr}$. 

The conformal structure of the quantum black hole is very similar to that of the classical 
\RN spacetime. In particular its ($r=0$)-hypersurface is timelike, in contradistinction 
to the Schwarzschild case where it is spacelike. In this respect the classical limit 
$\hbar\rightarrow 0$ is discontinuous, as is the limit $e\rightarrow 0$ of the \RN black hole. 

The Hawking temperature of very heavy quantum black holes is given by the semiclassical
$1/M$-law. As $M$ decreases, $T_{\rm BH}$ reaches a maximum at 
$\widetilde{M}_{\rm cr}\approx 1.27 M_{\rm cr}$ and then drops to $T_{\rm BH}=0$ at 
$M=M_{\rm cr}$. The specific heat capacity has a singularity at $\widetilde{M}_{\rm cr}$.
It is negative for $M>\widetilde{M}_{\rm cr}$, but positive for  
$\widetilde{M}_{\rm cr}>M>M_{\rm cr}$. We argued that the vanishing temperature of the critical 
black hole leads to a termination of the evaporation process once the black hole has reduced its mass
to $M=M_{\rm cr}$. This supports the idea of a cold, Planck size remnant as the final state of the evaporation.
For an infinitely far away static observer this final state is reached after an {\it infinite}
time only. 

For $M>M_{\rm cr}$, the entropy of the quantum black hole is a well defined, monotonically 
increasing function of the mass. For heavy black holes we recover the classical expression
${\cal A}/4 G_0$. The leading quantum corrections are proportional to $\ln(M/M_{\rm cr})$. 

In conclusion we believe that the idea of the renormalization group improvement which,
in elementary particle physics, is well known already is a promising new tool in order to 
study the influence of quantized gravity on the structure of spacetime. In the present work 
we focused on black holes, but it is clear that this approach has many more potential 
applications such as the very early universe, for instance. 

\section*{Acknowledgements}
One of us (M.R.) would like to thank the Department of Physics of Catania University for the cordial 
hospitality extended to him while this work was in progress. 
He also acknowledges a NATO traveling grant. A.B. would like to thank the 
Department of Physics of Mainz University for  financial support and for the cordial 
hospitality extended to him when this work was in progress.  
We are also grateful to INFN, Sezione di Catania for financial support.

\appendix{}
\renewcommand{\theequation}{A.\arabic{equation}}
\setcounter{equation}{0}
\section{The statistical mechanical entropy}
Our previous computation of $S(M)$ in section VII is within the framework of ``phenomenological'' 
thermodynamics. Ultimately one would like to derive this thermodynamics from the statistical 
mechanics based upon a fundamental Hamiltonian $\hat{H}$ which describes the microscopic 
degrees of freedom of both gravity and matter. The aim would be to compute a partition function
like $Z(\beta)={\rm Tr}[\exp (-\beta \hat{H})]$ and then to derive the free energy $F$, the 
internal energy $U$, the entropy $S$ and similar thermodynamic quantities from it
$(T\equiv 1/\beta)$:
\ba\label{7.7}
&&F=-T\;{\rm ln} Z\\[2mm]
&&U=T^2 {\partial \over \partial T}\;{\ln}Z\label{7.8}\\[2mm]
&&S=-{\partial F\over \partial T}\label{7.9}
\ea

In the original work of Gibbons and Hawking \cite{gh} the partition function was taken 
to be the functional integral of the pure Euclidean quantum gravity, 
$Z(\beta)=\int {\cal D}g_{\mu\nu} \exp (-I[g])$, where the integration is over all Euclidean 
metrics which are time periodic with period $\beta$. (Here $I[g]$ denotes the 
Einstein-Hilbert action with the Gibbons-Hawking surface term included.) The saddle point 
approximation of the integral yields, to leading order, 
\be\label{7.10}
Z(\beta)\approx \sum_{g_0^{\rm class}} e^{-I[g_0^{\rm class}]}
\ee
where the ``sum'' is over all saddle points $g_0^{\rm class}$ of $I$ with period $\beta$.
Considering only saddle points of the Schwarzschild black hole type, the latter requirement
means that only the hole of mass $M=\beta /8\pi G_0$ is relevant. For this
``Gibbons-Hawking instanton'', $\beta_{\rm BH}$ equals the externally prescribed value of 
$\beta$ (``on-shell'' approach). By using its action $I=4\pi G_0 M^2$ in 
$-\ln Z = \beta F \approx I$ one can derive the entire classical black hole thermodynamics.

It seems plausible to assume that the exact quantum gravity partition function 
in the Schwarzschild black hole sector is of the form
\be\label{7.11}
Z(\beta) = e^{-\Gamma [ g_0]}  
\ee
where $\Gamma[g]$ is some effective action functional, and $g_0$ is a stationary point
of $\Gamma$ with the same topology as $g_0^{\rm class}$:
\be\label{7.12}
{\delta \Gamma \over \delta g_{\mu\nu}} [ g_0] = 0
\ee
$\Gamma$ and $g_0$ are the quantum corrected versions of $I$ 
and $g_0^{\rm class}$, respectively. We set 
\be\label{7.13}
\Gamma = I +\Gamma_{\rm quant}
\ee
so that $\Gamma_{\rm quant}$ encapsulates the quantum effects. (The statistical mechanics  
based upon the one-loop approximation 
$\Gamma_{\rm quant}={1\over 2}{\rm ln}\;{\rm det}(\delta^2 I/ \delta g^2)$ has already been developed
to some extent \cite{frol2}.) The partition function (\ref{7.11}) and the thermodynamics derived from
it contain quantum corrections of two types:
\begin{description}
\item{{\it (i)}} The saddle point $g_0$, the metric of the ``quantum black hole'', differs from the
classical instanton $g_0^{\rm class}$. 
\item{{\it (ii)}} In order to obtain $\beta F$, the metric $g_0$ is inserted into
$\Gamma$ rather than $I$. 
\end{description}

Coming back to the renormalization group approach, it is natural to identify the saddle 
point $g_0$ with the Euclidean version of the renormalization  group improved Schwarzschild metric, 
eq.(\ref{6.4}) with (\ref{4.4}), which is denoted $g_{\rm imp}$ from now on. In this manner the
quantum effects of $(i)$ are approximately taken into account. However, $g_{\rm imp}$ was obtained
by a direct improvement of a classical {\it solution} rather than of the classical {\it action}.
Thus, within the framework used in the present paper, we do not know the functional 
$\Gamma$ for which $g_{\rm imp}$ is an (approximate)  saddle point and which would 
determine the partition function via (\ref{7.11}). The best we can do in this situation is to 
tentatively neglect the quantum effects of $(ii)$, {\it i.e.} to assume that 
$\Gamma_{\rm quant}[g_{\rm imp}]$ is much less important than $I[g_{\rm imp}]$ 
and to approximate (\ref{7.11}) by
\be\label{7.14}
Z(\beta)\approx e^{-I[g_{\rm imp}]}
\ee

In the following we investigate if (\ref{7.14}) can give rise to an acceptable thermodynamics. 
We shall employ the ``off-shell'' formalism (conical singularity method) developed 
in ref.\cite{solo} to which we refer for further details.

We evaluate the action $I$ for a general Euclidean metric of the type 
(\ref{6.1}) or ({\ref{6.4}) where $f(r)$ is arbitrary to a large extent. We only assume that 
it has a simple zero at some $r_{+}$. Its asymptotic behavior is required to be
\be\label{7.15}
f(r)=1-{2 G_0 M\over r}+O({1\over r^2})
\ee
for some fixed constant $M$. Furthermore, we assume that the Euclidean time $\tau$ in $(\ref{6.4})$
is an angle-like, periodic variable with period $\beta$. Here $\beta$ is the argument of the 
partition function. It has a prescribed value which in general does not coincide with 
$\beta_{\rm BH}\equiv 4\pi / f'(r_{+})$. The corresponding Euclidean manifold is denoted 
${\cal M}_\beta$. 

If we introduce the $2\pi$-periodic rescaled time variable
\be\label{7.16}
\hat{\tau}\equiv {2\pi\over \beta}\; \tau
\ee
then, near the horizon, the metric (\ref{6.4}) becomes 
\be\label{7.17}
ds_E^2\approx \rho^2 \;\left ({\beta \over \beta_{\rm BH}}\right )^2\;d\hat{\tau}^2 +d\rho^2
+r_{+}^2 \;d\Omega^2 
\ee
which coincides with (\ref{6.5}) only for the ``on-shell'' value $\beta=\beta_{\rm BH}$.  
For $\beta\not=\beta_{\rm BH}$ the space ${\cal M}_\beta$ has a conical singularity 
at $\rho = 0$, the angular deficit being $\delta = 2\pi (1-\beta/\beta_{\rm BH})$.
As a consequence, the curvature scalar on
${\cal M}_\beta$ has a delta-function singularity at $\rho =0$. 

The Einstein-Hilbert action $I\equiv I_{\rm reg}+I_{\rm sing}$ on ${\cal M}_\beta$
consists of a regular part and a singular part containing the contribution from the 
delta-function singularity. The regular part $I_{\rm reg}\equiv I_V+I_S$
has a volume and a surface contribution, 
\ba\label{7.18}
&&I_V= -{1\over 16\pi G_0}\int_{{\cal M}_\beta} d^4x \; \sqrt{g} R\\[2mm]
&&I_S= -{1\over 8\pi G_0}\int_{\partial {\cal M}_\beta} d^3x\; \sqrt{\gamma}
(K-K_0)\label{7.19}
\ea
where $\gamma$ and $K$ are the metric and the extrinsic curvature on the boundary
$\partial {\cal M}_\beta$ at infinity $(r\rightarrow\infty)$. ($K_0$ is the corresponding 
value for a flat metric.) 

The volume contribution is evaluated most easily by returning to the original $r$-coordinate.
Then, after performing the trivial angle and $\tau$-integrations, 
\be\label{7.20}
I_V=-{\beta\over 4 G_0}\int_{r_{+}}^\infty dr \; r^2 R(r)
\ee
where $R$ is the curvature scalar for the metric (\ref{6.1}). It reads 
\be\label{7.21}
R(r) = -{1\over r^2}\; \left [ {d^2\over dr^2}\Big (r^2 f(r) \Big )-2\right ]
\ee
and therefore the integral (\ref{7.20}) feels the behavior of $f$ only at the horizon
and at infinity:
\be\label{7.22}
I_V = \beta\;{r_{+}\over 2G_0}-{\beta \over 4 G_0 }\;
r^2_{+}\; f'(r_{+}) -{1\over 2}\; \beta M
\ee
The evaluation of (\ref{7.19}) with (\ref{7.15}) proceeds as in the standard case 
\cite{gh}:
\be\label{7.23}
I_S={1\over 2}\beta M
\ee
Adding (\ref{7.23}) to (\ref{7.22}) cancels precisely the last term of (\ref{7.22}) 
which originated from the upper limit of the integral (\ref{7.20}). Using (\ref{6.3}) the sum
contains only data related to the horizon, 
\be\label{7.24}
I_{\rm reg}=\beta \; {r_{+}\over 2 G_0}-{\beta\over \beta_{\rm BH}}\; {{\cal A}\over 4G_0}
\ee
where ${\cal A}\equiv 4\pi^2 r^2_{+}$ is its area.

For the metric (\ref{7.17}), the singular contribution 
\be\label{7.25}
I_{\rm sing}=-{1\over 16\pi G_0}\int_{{\cal M}_\beta}d^4x\;\sqrt{g}
R_{\rm sing}
\ee
with $R_{\rm sing}\propto \delta (\rho)$ has already been evaluated in ref.\cite{solo2}.
The result is
\be\label{7.26}
I_{\rm sing}=-\left (1-{\beta\over \beta_{\rm BH}}\right ) {{\cal A}\over 4 G_0}
\ee
Adding (\ref{7.26}) to (\ref{7.24}) we obtain the complete action 
evaluated on ${\cal M}_\beta$:
\be\label{7.27}
I=\beta\; {r_{+}\over 2 G_0}-{{\cal A}\over 4 G_0}
\ee
This is the result we wanted to derive. We emphasize that it is valid for black holes with 
an essentially arbitrary $f(r)$ and, as a consequence, arbitrary ADM-mass $M$ and Hawking temperature
$T_{\rm BH}=\beta_{\rm BH}^{-1}$. 

If we specialize for the renormalization group improved Schwarzschild black hole of a given
mass $M$, the action becomes 
\be\label{7.28}
I[g_{\rm imp}]=\beta \; {r_{+}(M)\over 2 G_0}-{{\cal A}(M)\over 4 G_0}
\ee
with $r_{+}(M)$ and ${\cal A}(M)$ given by (\ref{4.27}) and (\ref{6.26}), 
respectively.

If we tentatively insert the action (\ref{7.28}) into (\ref{7.14}) and use (\ref{7.9}) 
to calculate the entropy from $F\approx \beta^{-1} I[g_{\rm imp}]$ we obtain
\be\label{7.29}
S={{\cal A}(M) \over 4 G_0}
\ee
Apart from the modified relation between ${\cal A}$ and $M$, this is precisely the classical 
entropy. It is clear from eq.(\ref{6.26}) that (\ref{7.29}) differs from the correct
result (\ref{7.4}) already at the leading order of the large-$M$ corrections. 

Thus we must conclude that the ``statistical mechanics entropy'' (\ref{7.29}) fails to reproduce the
quantum corrections contained in the ``thermodynamical entropy'' (\ref{7.4}). The lesson to be learnt 
from this failure is that, at least as far as the entropy is concerned, the quantum mechanical
modification of the action from $I$ to $\Gamma$ is essential.
Improving only the saddle point $(g^{\rm class}_0\rightarrow g_{\rm imp})$ but 
neglecting $\Gamma_{\rm quant}$ is not sufficient in order to obtain a meaningful partition function.



\begin{references}
\bibitem{wilson} K.Wilson, J.Kogut, Phys. Rep. 12C, 75,  (1974).
\bibitem{rovelli}For a recent review see: C.Rovelli,
Contribution to the Special Issue 2000 of the Journal of Mathematical Physics, 
e-Print Archive: hep-th/9910131.
\bibitem{polc}J.Polchinski, Nucl.Phys. B231, 269, (1984).
\bibitem{ringwald} A.Ringwald, C.Wetterich, Nucl.Phys. B334, 
506, (1990); C.Wetterich, Nucl.Phys. B352, 529, (1991);
S. Liao, J.Polonyi, Ann. Phys. 222, 122, (1993);
M.Reuter, C.Wetterich, Nucl.Phys. B391, 147, (1993); Nucl.Phys. B408, 91, (1993).
\bibitem{abo} A.Bonanno, Phys. Rev. D52, 969, (1995);
A. Bonanno, D. Zappal\`a, Phys. Rev. D55, 6135, (1997) 
and hep-ph/9611271.
\bibitem{wet}  C.Wetterich, Phys.Lett. B301, 90, (1993);
M.Reuter, C.Wetterich, Nucl. Phys. B417, 181, (1994), 
Nucl. Phys. B427, 291, (1994), Nucl.Phys. B506, 483, (1997). 
\bibitem{reuter} M.Reuter, Phys. Rev. D57, 971, (1998), 
e-Print Archive: hep-th/9605030.
\bibitem{ue}W.Dittrich, M.Reuter, {\it Effective Lagrangians in 
Quantum Electrodynamics, } Springer, Berlin, 1985.
\bibitem{bore} A.Bonanno, M.Reuter, Phys.Rev. D60, 084011, (1999), 
e-Print Archive: gr-qc/9811026. 
\bibitem{corfu} For an introduction see: M.Reuter in
{\it Proceedings of the 5th Hellenic School and Workshop
on Elementary Particle Physics}, Corfu, Greece, (1995), e-print Archive:
hep-th/9602012. 
\bibitem{we} S.Weinberg in {\it General Relativity: an Einstein
Centenary Survey}, S.W.Hawking and W.Israel (Eds.), Cambridge 
University Press, (1979).
\bibitem{souma} W.Souma, Prog. Theor. Phys. 102, 181, (1999), e-print 
Archive: hep-th/9907027.
\bibitem{pol} A.Polyakov in {\it Gravitation and Quantization}, Proceedings of the 
Les Houches Summer School, Les Houches, France, 1992, J.Zinn-Justin, B.Juli\`a 
(Eds.), Les Houches Summer School Proceedings Vol. 57, North-Holland, Amsterdam, (1995),
e-Print Archive: hep-th/9304146. 
\bibitem{don}J.F.Donoghue, Phys. Rev. Lett. 72, 2996, (1994);
Phys. Rev. D50, 3874, (1994).
\bibitem{MTW} C.W.Misner, K.S.Thorne, J.A.Wheeler, {\it Gravitation}, 
W.H.Freeman \& Co., San Francisco (1973).
\bibitem{hamber} H.W.Hamber, S.Liu, Phys. Lett. B357, 51, (1995).
\bibitem{frol}V.P.Frolov, M.A. Markov, V.F. Mukhanov, Phys.Rev.D41, 383, (1990). 
\bibitem{dimnikova} I.Dymnikova, Gen.Rel.Grav. 24, 235, (1992).
\bibitem{strominger} A. Strominger, Phys. Rev. D46, 4396, (1992).  
\bibitem{hasslacher} B. Hasslacher, E. Mottola, Phys.Lett. B99, 221, (1981). 
\bibitem{euclidean} S.W.Hawking in {\it General Relativity: an Einstein
Centenary Survey}, S.W.Hawking and W.Israel (Eds.), Cambridge 
University Press, (1979).
\bibitem{gh} G.Gibbons, S.Hawking, Phys.Rev. D15, 2752, (1977).
\bibitem{wald} For an introduction see: 
R.M.Wald, {\it General relativity}, University of Chicago Press,
(Chicago, 1984). 
\bibitem{frol2} V.P.Frolov, W.Israel, S.N.Solodukhin, Phys.Rev. D54, 2732, (1996). 
\bibitem{solo}S.N. Solodukhin, Phys.Rev. D54, 3900, (1996). 
\bibitem{solo2}R.B. Mann, S.N. Solodukhin, Phys.Rev. D54, 3932, (1996).
\end{references}
\end{document}